\documentclass[reprint,twocolumn,amsmath,amssymb,aps,pra,longbibliography,floatfix]{revtex4-1}
\usepackage{physics}
\usepackage{amsmath}
\usepackage{cases}
\usepackage{url}
\usepackage{natbib}
\usepackage{textcase}
\usepackage{amssymb}
\usepackage{graphicx}
\usepackage{bm} 
\usepackage{times}
\usepackage{float}
\usepackage{multirow,microtype,color,relsize}
\usepackage{subfigure}
\usepackage[breaklinks=true]{hyperref}
\usepackage[utf8]{inputenc}
\usepackage[english]{babel}
\hypersetup{colorlinks=true,linkcolor=blue,citecolor=blue}
\hypersetup{linktocpage}
\usepackage{CJKutf8}
\usepackage{breakcites}
\usepackage{float}
\usepackage[dvipsnames]{xcolor}
\definecolor{mypine}{RGB}{1, 121, 111}

\begin{document}
\title{Electronic transport, metal-insulator transition, and Wigner crystallization in transition metal dichalcogenide monolayers}
\author{Yi Huang}
\author{Sankar Das Sarma}
%\email[Corresponding author: ]{huan1756@umn.edu}
\affiliation{Condensed Matter Theory Center and Joint Quantum Institute, Department of Physics, University of Maryland, College Park, Maryland 20742, USA}
\begin{abstract}	
Two recent electronic transport experiments from Columbia University and Harvard University have reported record high mobility and low channel densities in flux-grown transition metal dichalcogenide (TMD) WSe$_2$ monolayers [J. Pack, 
\textit{et al.}, \href{https://arxiv.org/abs/2310.19782}{arXiv:2310.19782} [cond-mat.mes-hall]; A. Y. Joe, \textit{et al.}, \href{https://doi.org/10.1103/PhysRevLett.132.056303}{Phys. Rev. Lett. \textbf{132}, 056303 (2024)}]. 
A two-dimensional (2D) metal-insulator transition (MIT) is demonstrated in the Columbia sample at low densities, a regime where the formation of a Wigner crystal (WC) is theoretically anticipated in the absence of disorder. We employ the finite-temperature Boltzmann theory to understand the low-temperature transport properties of monolayer TMDs, taking into account realistic disorder scattering. We analyze the experimental results, focusing on the 2D MIT behavior and the influence of temperature and density on mobility and resistivity in the metallic phase. We provide a discussion of the nontrivial carrier density dependence of our transport results. Our analysis elucidates the linear temperature-dependent resistivity observed in the metallic phase, attributing it to Friedel oscillations associated with screened charged impurities. Furthermore, we explore whether Coulomb disorder could lead to the MIT through either a quantum Anderson localization transition or a classical percolation transition in the long-range disorder potential landscape. Our theoretical estimates of the disorder-induced MIT critical densities, although smaller, are within a factor of approximately 2 of the experimental critical density. We examine the exceptionally high melting temperature $\sim$10 K of WCs observed experimentally in the MoSe$_2$ systems at low density, which is strongly enhanced by the disorder-induced localization effect, since the pristine melting temperature is an order of magnitude smaller. This suggests that the observed 2D low-density MIT behavior is likely a result of the complex interplay between disorder effects and interaction-driven WC physics, offering a comprehensive understanding of the low-temperature transport phenomena in TMD monolayers.
\end{abstract}
\maketitle
%\onecolumngrid
\section{Introduction}\label{sec:introduction}
The atomically thin two-dimensional (2D) transition metal dichalcogenides (TMDs) hold great promise for future spintronics, valleytronics, and optoelectronics~\cite{Mak_KinFai:2010,Radisavljevic:2011,Korn:2011,Avouris_Heinz_Low_2017} due to their strong spin-valley coupling~\cite{Xiao:2012,Xu_Xiaodong:2014}, for which high-mobility samples with long spin and valley lifetimes are essential.
More crucially for our purpose, they are also strongly interacting continuum electron systems with large effective mass and relatively low lattice dielectric constant enhancing the inter-particle Coulomb coupling.
A wide range of interaction-induced phenomena have been observed in TMD monolayers and moir\'{e} structures, including Wigner crystals (WCs)~\cite{Smolenski:2021,sung2023observation,xiang2024quantum}, fractional quantum Hall states~\cite{Shi:2020,CDean_WSe2:2023}, integer and fractional Chern insulators (quantization of anomalous Hall effect due to band topology, without any Landau quantization imposed by an external magnetic field)~\cite{Li_Tingxin:2021,foutty2023mapping,Cai:2023a,Zeng:2023,Park:2023,Li_Tingxin:2023,Zhao:2024}, Kondo lattice~\cite{Wenjin_Zhao:2023}, and exciton insulators~\cite{Ma:2021,nguyen2023perfect,qi2023perfect}.
Despite these advances, the presence of unintentional Coulomb disorder in TMDs arising from random quenched charged impurities poses significant challenges, influencing the transport properties and the manifestation of many interaction-driven phenomena~\cite{Ahn_MIT_mTMD:2022,Ahn_temperature_mTMD:2022,Poduval_Kondo:2022,sarma2024zerofield}. Understanding and controlling disorder is crucial for enhancing the purity of TMD samples, thereby unlocking their full potential for both technological applications and fundamental research. 
Recent advances in synthesis techniques and device fabrication have led to cleaner TMD samples, showcasing record-high mobilities and lower defect densities, which are essential for probing new quantum phenomena~\cite{liu2023twostep,PKim:2023,CDean_WSe2:2023}.
For example, scanning tunneling microscopy (STM) studies have shown that a recently developed flux growth technique reduces the density of charged-neutral point defects in WSe$_2$ from $10^{13}$ cm$^{-2}$ to below $10^{11}$ cm$^{-2}$~\cite{liu2023twostep}.
In this context, recent experiments from Columbia~\cite{CDean_WSe2:2023} and Harvard University~\cite{PKim:2023} have demonstrated unprecedented mobilities in flux-grown WSe2 monolayers, alongside the ability to achieve low channel carrier densities. 
The Columbia sample shows a record high mobility $\sim 8\times10^4$ cm$^{2}$/Vs at a temperature $T=1.5$ K, and the Harvard sample shows a comparable mobility of around $2.5 \times 10^4$ cm$^{2}$/Vs at $T=1.7$ K.
By contrast, typical mobilities in older TMD samples are $\sim10^3$ cm$^2$/Vs.
Using transparent Ohmic contacts, it is possible to reach a channel carrier density as low as $1.5 \times 10^{11}$ cm$^{-2}$~\cite{CDean_WSe2:2023}, which is an order of magnitude lower than those reported in past generations of devices.
%The flux-grown WSe$_2$ monolayer exhibits record-high hole mobility, $\sim 8\times10^4$ cm$^{2}$/Vs at a low temperature $1.5$ K and $\sim 2\times10^3$ cm$^{2}$/Vs at room temperature, as reported by a group from Columbia University~\cite{CDean_WSe2:2023}.
%Experimentalists at Harvard University have shown a comparable high mobility of around $2.5 \times 10^4$ cm$^{2}$/Vs at a temperature of $1.7$ K~\cite{PKim:2023}.
These advances have facilitated the putative observation of a 2D metal-insulator transition (MIT) in a density regime near $r_s \sim 30$~\cite{CDean_WSe2:2023} where the WC formation is theoretically expected in the absence of disorder~\cite{Drummond:2009}.
%They also showed, in any van der Waals material beyond graphene, the first transport signature of fully developed fractional quantum Hall states in monolayer WSe$_2$~\cite{CDean_WSe2:2023}.
Here, the dimensionless interaction parameter $r_s = 1/\sqrt{\pi n a_B^2}$ represents the ratio of the Coulomb interaction to the kinetic energy, where $n$ is the 2D carrier density and $a_B = \kappa \hbar^2/m e^2$ is the effective Bohr radius.
(In the rest of this paper, we interchangeably refer to $n$ as $n_h$ and $n_e$ for hole and electron charge carriers, respectively.)
%The Columbia group claimed the nature of the 2D MIT is likely an interaction-driven Wigner crystal transition due to a relatively large value of the critical $r_s \approx 25$ where the conductivity vanishes experimentally.
The magneto-optical measurement on monolayer MoSe$_2$ reported by the Harvard group also shows evidence of density-tuned WC ($r_s >30$) and the microemulsion mixed state ($r_s\approx 20-30$) where regions of WCs and Fermi liquid coexist across multiple length scales~\cite{sung2023observation}.
Similarly to GaAs hole~\cite{Shayegan:1999,Shayegan:1991,Pfeiffer:2018,Shayegan:2020}, AlAs~\cite{Hossain:2020,Hossain:2022} and ZnO~\cite{Falson2022} electron systems where WCs are observed, the large effective mass $m\sim 0.5 m_0$ in TMD monolayers suppresses the kinetic energy relative to the Coulomb energy, where $m_0$ is the free electron mass.
In addition, the monolayer WSe$_2$ displays high mobility over a large range of $r_s$, similar to AlAs and ZnO devices, and exceeding the range of $r_s$ values accessible in Si MOSFET devices by more than a factor of 2, creating the opportunity to observe WCs in TMD materials.
However, it is not clear whether disorder plays an important role so that the observed MIT is, instead of the interaction-driven WC formation, actually a quantum Anderson localization transition or a classical percolation transition occurring accidentally at large $r_s$ (see Ref.~\cite{Ahn_Anderson_Wigner:2023} for the discussion in other 2D systems).
In this context, we mention that the extensively-studied 2D n-GaAs systems, although having much higher mobility $10^7$ cm$^2$/Vs, typically have much smaller $r_s$ ($<$15) values even at an electron density as low as $10^9$ cm$^{-2}$~\cite{Lilly:2003}. 
%Therefore, it is timely to study the disorder effect in the flux-grown high-mobility TMD samples.
\begin{figure}[t]
    \centering
    \includegraphics[width = 0.8\linewidth]{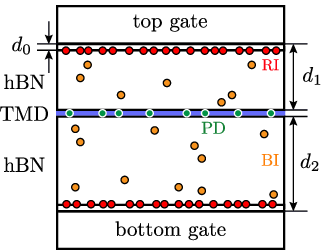}%mu_low_ne0.eps
    \caption{A schematic illustration of the high-mobility monolayer TMD reported in Refs.~\cite{CDean_WSe2:2023,PKim:2023}. The red, orange, green circles represent the delta-layer remote charged impurities (RI), uniform background charged impurities (BI), and charge-neutral atomic point defects (PD) respectively.}
    \label{fig:schematics}
\end{figure}

In this paper, we develop a theory for electronic transport in TMD monolayers with realistic disorder, explaining the experimental resistivity data and enhancing our quantitative understanding of disorder effects in TMDs, which is essential for further improvement of the sample mobility in the future.
%As a reminiscence of the history of the discovery of even denorminator FQHE in GaAs due to a huge improvement in the mobility, 
%Understanding disorder effects paves the way for further improvement of sample mobility in the future, which is essential for discovering new interaction-driven phenomena on the TMD platform.
The main goal of this paper is to study the transport scattering mechanisms due to screened charged impurities and charged-neutral point defects in the continuum (i.e. non-moir{\'e}) flux-grown TMD monolayers at low temperatures within the effective mass approximation and the finite-temperature Boltzmann transport theory.
See Fig.~\ref{fig:schematics} for the device structure. 
More specifically, we explain the observed linear-in-$T$ resistivity in the metallic phase, attributed to the temperature-dependent Friedel oscillations from screened charged impurities~\cite{Hwang_low_temp:2015,Zala_Aleiner:2001}.
(This physics could qualitatively be understood as the analog of the 2D Altshuler-Aronov effect at temperatures $T\tau/\hbar \gtrsim 1$~\cite{Zala_Aleiner:2001,Altshuler_Aronov:1985}, where $\tau$ is the momentum relaxation time.)
The origin of these Friedel oscillations in real space can be traced back to the kink in the screening polarizability function at the backscattering momentum, denoted as $q=2k_F$~\cite{Stern:1967,Das_Sarma_screening:1986}. 
Notably, the effect of temperature in diminishing the $2k_F$ screening is profound in 2D, following a $T^{1/2}$ trend, in contrast to the exponentially weak temperature dependence observed in the long wavelength limit $q\to0$. 
At low temperatures, the coherent interference of electrons scattered from the Friedel oscillations surrounding a charged impurity enhances the backscattering with a correction to the scattering cross section proportional to $T/E_F$. 
Here, $E_F$ is the Fermi energy and $k_F$ is the Fermi wave vector.
This low-temperature correction eventually leads to the linear-in-$T$ resistivity, which violates the well-known Sommerfeld expansion, where the leading correction should be quadratic.
We emphasize that the metallic resistivity in our theory and presumably in the low-temperature data reported in Ref.~\cite{CDean_WSe2:2023}, which depends linearly on temperature $T$, does not originate from any quantum criticality at $T = 0$ and does not suggest any non-Fermi-liquid characteristics.
It also does not arise from phonon scattering since the temperature of interest ($T\lesssim 10$ K) is in the low-$T$ Bloch-Gruneisen (BG) regime where phonon scattering is suppressed~\cite{Poduval_Kondo:2022}. 
%It is also important to note that any temperature variation resulting from phonon scattering is insignificant at low temperatures of interest $T\lesssim 10$ K, due to the weak electron-phonon coupling in TMDs.
In this work, we explore the possibility that this same Coulomb disorder is also responsible for producing the 2D MIT at low carrier densities as (1) a strong disorder-driven quantum Anderson localization transition or (2) a classical percolation transition through the long-range disorder potential landscape.
For the first scenario, using the Anderson-Ioffe-Regel (AIR) condition $k_F l = 1$ or equivalently $\rho = h/e^2$~\cite{anderson1958,ioffe1960}, we compute the effective MIT critical density $n_c$ and find that the apparent $n_c$ increases as temperature increases, because the effective Coulomb disorder is enhanced with increasing temperature leading to reduced screening.
Here, $l = v_F \tau$ is the transport mean free path and $\rho = m/e^2 n \tau$ denotes the resistivity.
The low-$T$ theoretical AIR critical density $n_c \approx 6\times10^{10}$ cm$^{-2}$ agrees with the estimation reported in Ref.~\cite{CDean_WSe2:2023}, where $n_c$ is estimated by setting $E_F$ equal to the disorder broadening $\Gamma_q = \hbar/2\tau_q$ with the single-particle relaxation time $\tau_q \approx 1$ ps measured by the Shubunikov-de Hass oscillations. 
However, this theoretical AIR critical density is somewhat lower than the experimental critical density $\sim 1.5 \times 10^{11}$ cm$^{-2}$ measured at $T= 1.5$ K. 
The experimental $n_c$ is defined as the density where the resistivity slope $d\rho/dT$ changes sign at low temperatures, whereas the AIR criterion for $n_c$ is essentially a $T=0$ theoretical criterion.
This discrepancy could be attributed to four potential reasons: first, the breakdown of the Boltzmann theory at low densities in the presence of Coulomb impurities; second, the possibility that quantum Anderson localization is replaced by a classical percolation crossover at finite temperatures; third, the emergence of an interaction-driven WC transition at large $r_s$ value; fourth, our screening theory based on the random phase approximation (RPA) becomes increasingly inaccurate at lower densities where the enhanced $r_s$ introduces strong correlation effects (eventually leading to Wigner crystallization) neglected in the RPA screening.
%given little disorder and the experimental critical $r_s$ is close to the theoretical critical $r_s = 30$ computed by quantum Monte Carlo simulation~\cite{Drummond:2009}.
We elaborate on the percolation scenario in the presence of long-range Coulomb disorder as follows. 
The 2D system at low densities is broken into inhomogeneous charge puddles separated by long-range Coulomb disorder potential barriers~\cite{Thouless:1971,Shklovskii:1972,Kirkpatrick:1973,Shklovskii:1975,Ando_review:1982,shklovskii1984,DasSarma:2005579,Shklovskii:2007,Shaffique:2007,Manfra:2007,Tracy:2009,DasSarma:2005,Qiuzi:2013,Tracy:2014}, so the Boltzmann transport theory with linear screening is no longer applicable.
This scenario supports a classical percolation model, where metallic electron puddles navigate through a disordered potential landscape. 
We note that at $T=0$ quantum tunneling among the random puddles becomes important, making the classical percolation scenario equivalent to the Anderson localization phenomenon as a matter of principle, but physically, they are quite different as Anderson localization arises from quantum interference whereas the percolation transition arises from the nonexistence of a conducting path through a disorder landscape.
These two localization scenarios may happen at low densities or high $r_s$, which could be close to the theoretical WC values, particularly in clean systems~\cite{Ahn_Anderson_Wigner:2023}.
By analyzing low-density data in WSe$_2$ monolayers~\cite{CDean_WSe2:2023}, we identify a percolation threshold density $n_p\approx 1\times10^{11}$ cm$^{-2}$ ($r_s \approx 30$), which, while higher than the AIR critical density $n_c\approx 6\times10^{10}$ cm$^{-2}$ ($r_s \approx 40$), remains below the experimental critical density $n_c\approx 1.5 \times 10^{10}$ cm$^{-2}$ ($r_s \approx 25$).
All of them have high values of $r_s$ close to the theoretical WC transition predicted by the quantum Monte Carlo (QMC) calculation in the absence of disorder, where $r_s \approx 30$~\cite{Drummond:2009}.
These findings suggest that the observed 2D MIT behavior likely results from the complex interplay between disorder effects and interaction-driven WC physics.
%Therefore, the underlying physical picture of the observed 2D MIT should be the interplay between disorder and interaction, where disorder cannot be ignored~\cite{Ahn_Anderson_Wigner:2023}.
To explore the relationship between disorder and the 2D MIT, we also theoretically predict the AIR $n_c$ as a function of the maximum mobility $\mu_{\max}$ deep into the metallic state (the peak mobility achieved by tuning the carrier density at the lowest accessible temperature) by intentionally adding disorder and making dirtier samples, finding that $n_c$ increases as $\mu_{\max}$ decreases~\cite{Ahn_MIT_Si:2022}.
Here, $\mu_{\max}$ serves as an approximate experimental measure of the sample disorder-- the lower the disorder, the higher is $\mu_{\max}$.
This analysis emphasizes the significant role of disorder in the 2D MIT and sets the stage for future experimental investigations.
%To examine the perspective of WC in high mobility WSe$_2$ samples, future experiments are necessary, for example, depinning the WC by increasing the current, and measuring the spin susceptibility to show competing magnetic orders.
Moreover, we extend our theoretical predictions to the resistivity and mobility as functions of carrier densities for the monolayer MoSe$_2$ experiment conducted at Harvard University~\cite{sung2023observation}, despite the lack of direct transport measurement data.
In particular, we comment on the WC melting temperature $T_m\sim$ 10 K experimentally measured in the MoSe$_2$ systems~\cite{Smolenski:2021}, much higher than the mean-field theoretical prediction $T_m\sim$ 1 K for a pristine WC~\cite{Hwang_WC:2001}. 
This shows that the effective melting temperature is strongly enhanced by the disorder-induced localization effect~\cite{DinhDuy_Vu:2022}.
Nevertheless, the high melting temperature and large critical density of WC in TMDs, attributed to their large effective mass and small background lattice dielectric constant, make TMDs a suitable material platform for exploring the WC physics.

The structure of this paper is as follows.
Section~\ref{sec:main_results} summarizes the results of temperature-dependent resistivity and mobility in WSe$_2$ hole systems, alongside a detailed comparison with the experimental data from both the Columbia~\cite{CDean_WSe2:2023} and Harvard~\cite{PKim:2023} groups. 
Additionally, we discuss the role of disorder in the context of the magneto-optical experiments conducted by a different Harvard group on higher-disorder samples~\cite{sung2023observation}, though the transport experimental data are not available for these samples, making the theoretical analysis somewhat uncertain.
We discuss the disorder-enhanced melting temperature of WC in Sec.~\ref{sec:WC_melting}.
In Sec.~\ref{sec:review_polarizability}, we provide the technical details for the 2D polarizability and Friedel oscillations at finite temperatures. 
This section aims to elucidate the physical picture of the coherent interference of electrons scattered by Friedel oscillations, which leads to the observed linear-in-$T$ resistivity in the metallic phase arising from the screening of the impurity disorder.
Section~\ref{sec:t-dependent_resistivity} gives a more analytical exploration of the temperature-dependent resistivity, focusing on the effects of scattering by random charged impurities and charge-neutral atomic point defects.
Section~\ref{sec:gate} examines the screening effect caused by the double-gate configuration, commonly used in transport experiments.
Section~\ref{sec:RI_BI} compares the RI and BI scattering at low densities and suggests a way to improve the low-density mobility by increasing the hBN thickness in experiments.
Section~\ref{sec:phonon} comments on the effect of electron-phonon scattering at high temperatures.
Finally, we conclude and summarize in Sec.~\ref{sec:conclusion}.
%The conduction and valence band minima in monolayer TMDs are reached at the corners ($K$ points) of the first Brillouin zone, and away from the band minima the broken inversion symmetry combined with the strong spin-orbit coupling lifts the fourfold (spin-valley) degeneracy, and yields coupled spin and valley degrees of freedom~\cite{Xiao:2012}.
\begin{table}[b]
\caption{Material parameters used in the calculations. The effective mass $m$ is expressed in units of the free electron mass $m_0$. $\kappa$ is the dielectric constant. $g$ is the total quantum degeneracy.}
\begin{ruledtabular}
\begin{tabular}{c | c | c | c} 
 material & $m/m_0$ & $\kappa$ & $g$ \\
%heading
\hline \\ [-2ex] 
%\hline \\
WSe$_2$ hole & 0.45~\cite{Fallahazad:2016,Ahn_temperature_mTMD:2022,Ahn_MIT_mTMD:2022,CDean_WSe2:2023} & 5~\cite{Emanuel:2018,Laturia:2018,CDean_WSe2:2023} & 2~\cite{Xiao:2012,Xu_Xiaodong:2014} \\ 
MoSe$_2$ electron & 0.7~\cite{sung2023observation} & 4.6~\cite{sung2023observation} & 2~\cite{Xiao:2012,Xu_Xiaodong:2014} \\ 
\end{tabular}
\end{ruledtabular}
\vspace{-0.2 in}
\label{table:parameters}
\end{table}
\begin{figure}[t]
    \centering
    \includegraphics[width = \linewidth]{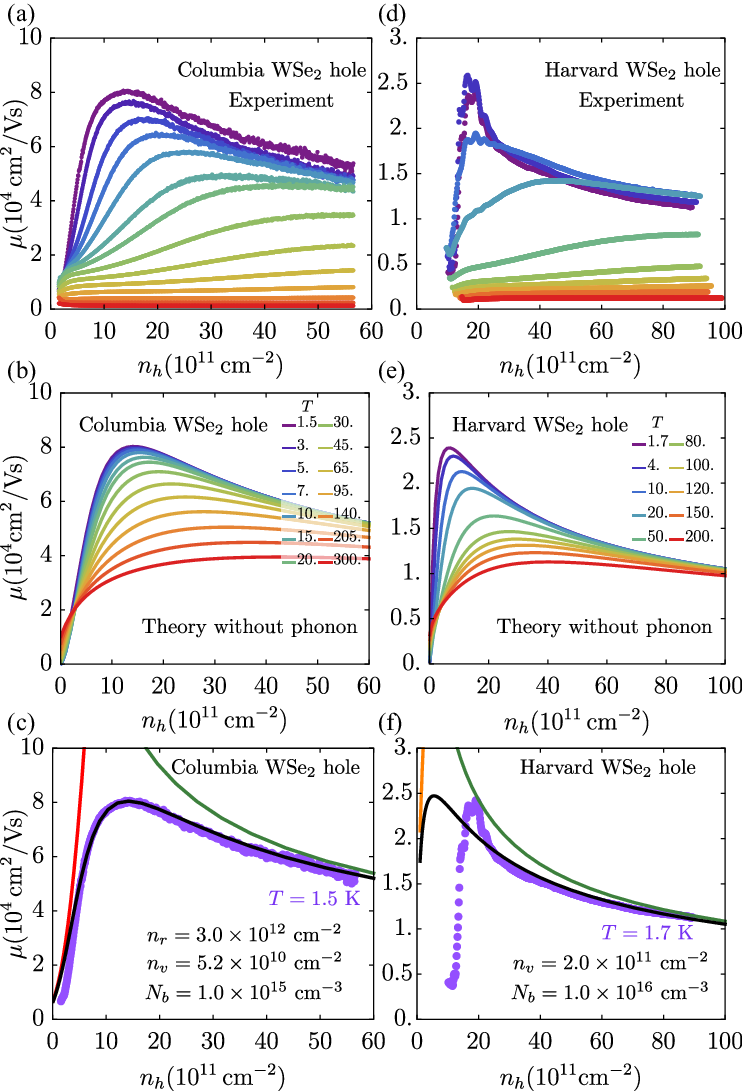}%mu_low_ne0.eps
    \caption{Mobility $\mu$ vs hole density $n_h$ in a monolayer WSe$_2$ sample from the (a) (b) (c) Columbia~\cite{CDean_WSe2:2023} and (d) (e) (f) Harvard~\cite{PKim:2023} groups. (a) and (c): Different colored dots represent experimental data measured at different temperatures. (b) and (e): The corresponding theoretical results are obtained by assuming only impurity scattering without phonon. The color legends of the temperature in units of K are shown as inserts. The impurity parameters are obtained from the best-fit of the low-$T$ mobility as shown in (c) and (f), where the solid black curves represent the theoretically calculated total mobility at $T=0$. The red, orange, and green curves are contributions from the remote charged impurity ($n_r$), background charged impurity ($N_b$), and charged-neutral point defect ($n_v$) scattering, respectively.}
    \label{fig:muT_n}
\end{figure}
\section{Main results}
\label{sec:main_results}
In this section, we present our main theoretical results and the comparison with experimental data of WSe$_2$ hole systems from the Columbia~\cite{CDean_WSe2:2023} and Harvard~\cite{PKim:2023} group.
We also comment on the MoSe$_2$ electron system studied by a different Harvard group~\cite{sung2023observation}, though there are no direct transport measurement data in this sample.
The transport mobility $\mu = e \tau / m$ is defined through the momentum relaxation time $\tau$, and the corresponding resistivity is given by the Drude formula $\rho = m/n e^2 \tau$.
The detailed calculation of the temperature dependent relaxation time can be found in Sec.~\ref{sec:t-dependent_resistivity}, with the focus being on the calculated results in this section.
%Although the details of the calculation will be discussed in Section~\ref{sec:t-dependent_resistivity}, we mention 
The material parameters we use in the calculations are summarized in Table~\ref{table:parameters}.
We use the effective mass $m=0.45m_0$ in WSe$_2$ hole systems~\cite{Fallahazad:2016,Ahn_temperature_mTMD:2022,Ahn_MIT_mTMD:2022,CDean_WSe2:2023} and $m=0.7m_0$ in MoSe$_2$ electron systems~\cite{sung2023observation}.
The total quantum degeneracy in monolayer TMDs is $g=2$ due to spin-valley locking~\cite{Xiao:2012,Xu_Xiaodong:2014}.
We use $\kappa \approx 5$ as the background lattice dielectric constant.
Mostly the contribution of the dielectric constant comes from the hBN environment~\cite{Emanuel:2018}, where the out-of-plane and in-plane dielectric constants are $\epsilon^{\perp} = 3.76$, $\epsilon^{\parallel} = 6.93$ respectively~\cite{Laturia:2018}, and $\kappa = \sqrt{\epsilon^{\perp} \epsilon^{\parallel}}$. 
We take the size of the atomic point defects as the lattice constant $a_0 = 3.32$ {\AA} for monolayer WSe$_2$~\cite{Junqiao:2013}.

Figure~\ref{fig:muT_n} shows both the experimental data and the calculated mobility $\mu$ as a function of hole density $n_h$ at various temperatures, where panels (a-c) represent the Columbia sample while panels (d-f) represent the Harvard sample.
At low temperatures $T\lesssim 20$ K, as depicted in Fig.~\ref{fig:muT_n} (a) and (d), the mobility $\mu(n_h)$ exhibits a non-monotonic behavior, peaking at a specific density $n_{\max}$.
For the Columbia sample, the maximum mobility, $\mu_{\max}$, is observed to be $8.0\times 10^4$ cm$^2$/Vs at a density of $n_{\max} = 1.5 \times 10^{12}$ cm$^{-2}$. 
The Harvard sample reaches a $\mu_{\max}$ of $2.5\times 10^4$ cm$^2$/Vs at a higher density of $n_{\max} = 2.0 \times 10^{12}$ cm$^{-2}$. 
These peak mobility values correspond to very long transport mean free paths of approximately $2$ $\mu$m and $0.5$ $\mu$m for the Columbia and Harvard samples, respectively.
As temperature increases, $\mu_{\max}$ decreases and $n_{\max}$ increases because of suppressed screening of the Coulomb disorder.
This trend is consistent across both samples and is similarly observed in commercial chemical vapor transport (CVT) crystals studied by the Harvard group~\cite{PKim:2023}.
However, the mobility in CVT crystals is an order of magnitude lower compared to the flux-grown samples, around $\sim$3,000 cm$^{2}$/Vs at 4 K. 
This comparison underscores the quality of the flux-grown samples and highlights the temperature-dependent behavior of mobility in these TMD systems.

The observed non-monotonic relationship between mobility and carrier density in TMD systems can be explained by a combination of two scattering mechanisms.
The increasing mobility as the density increases can be explained by the enhanced screening of the charged impurity scattering dominating at low densities $n < n_{\max}$. 
While at high densities $n > n_{\max}$, the decrease in mobility as density increases can be attributed to charge-neutral atomic point defects (PDs) like Se vacancies~\cite{PKim:2023}, since Coulomb disorder is already suppressed by screening at high densities.
As shown in Fig.~\ref{fig:muT_n}, our theory qualitatively and semiquantitatively agrees with the data. 
Our charge impurity model consists of delta-layer remote impurities (RIs) and a uniform distribution of background impurities (BIs). 
The delta-layer RIs, characterized by a 2D concentration $n_r$, are located near the interface between the hBN dielectric spacer and the gate, at a distance $z=d$ from the TMD monolayer. 
We choose $d$ such that the remote impurities are separated from the gates by a distance $d_0 \sim 1$ nm.
The reason for choosing RIs close to the gate is based on the empirical observation that transport data can be significantly affected by changing gate materials and assembly technology, particularly in van der Waals materials (see, for example, transport data in bilayer graphene systems~\cite{Icking:2022}). 
For the Columbia sample, the thickness of the top and bottom hBN gate dielectrics are $d_1 = 11$ nm and $d_2 = 25$ nm, respectively, whereas for the Harvard sample $d_1 = 50$ nm and $d_2 = 73$ nm.
The uniform distribution of BIs within the hBN dielectrics is characterized by a three-dimensional (3D) concentration $N_b$. 
Additionally, charge-neutral atomic point defects, with a concentration $n_v$, are assumed to be situated within the same plane as the TMD monolayer. 
See Fig.~\ref{fig:schematics} for a visual representation of the device structure and the spatial arrangement of these impurities. 

To estimate the impurity concentrations, in Figs.~\ref{fig:muT_n} (c) and (f), we perform a numerical fit to the mobility curves at the lowest temperature available in experiments $T\sim 1$ K.
From the best-fits we get for the Columbia sample $n_r=3.0\times 10^{12}$ cm$^{-2}$, $N_b=1.0\times 10^{15}$ cm$^{-3}$, and $n_v=5.2\times 10^{10}$ cm$^{-2}$; while for the Harvard sample $N_b=1.0\times 10^{16}$ cm$^{-3}$ and $n_v=2.0\times 10^{11}$ cm$^{-2}$.
Note that these numbers, although unknown in actual experimental samples, are reasonable for the TMD materials being studied here.
For example, these numbers are actually consistent with STM studies of the bulk TMD crystals~\cite{Edelberg:2019,liu2023twostep}.
We do not include the delta-layer RIs in the calculations for the Harvard sample for two reasons. 
First, the distances between the top and bottom gates are $d_1 = 50$ nm and $d_2 = 73$ nm, which are significantly larger than those in the Columbia sample. Consequently, the remote impurity layer is likely to be located at a greater distance from the TMD monolayer, resulting in a negligible Coulomb potential.
Second, remote impurity scattering manifests itself only at low densities $\lesssim 10^{12}$ cm$^{-2}$, where there is no reliable data due to large contact resistance for the Harvard sample. 
It should be noted that both experiments report a rapid increase in contact resistance below certain carrier densities, denoted as $n_{\mathrm{con}}$. 
Specifically, for the Columbia sample, $n_{\mathrm{con}}$ is identified at $1.7\times 10^{11}$ cm$^{-2}$ [cf. Fig. 1f in Ref.~\cite{CDean_WSe2:2023}], and for the Harvard sample, at $1.5\times 10^{12}$ cm$^{-2}$. 
%In the Columbia sample $n_{\mathrm{con}} = 1.7\times 10^{11}$ cm$^{-2}$ [cf. Fig. 1f in Ref.~\cite{CDean_WSe2:2023}], while in the Harvard sample $n_{\mathrm{con}} = 1.5\times 10^{12}$ cm$^{-2}$~\cite{PKim:2023}.
Given that the two-terminal resistance comprises both contact and bulk resistances, the bulk resistance data are more reliable at densities higher than $n_{\mathrm{con}}$.
The large value of $n_{\mathrm{con}}$ for the Harvard sample likely surpasses the critical density for observing a MIT, which accounts for the absence of MIT in their data.
In practice, the charged impurities may also be screened by the contact electrodes, whose effect is rather complicated and depends on the specific contact scheme and geometry. 
However, the essential physics of Coulomb impurity scattering should not be affected by contact screening, which should only change the result by a nominal numerical factor. For this reason, we do not consider the contact electrode screening, which would involve a complex calculation with many unknown parameters (rendering it essentially useless).
\begin{figure}[t]
    \centering
    \includegraphics[width = \linewidth]{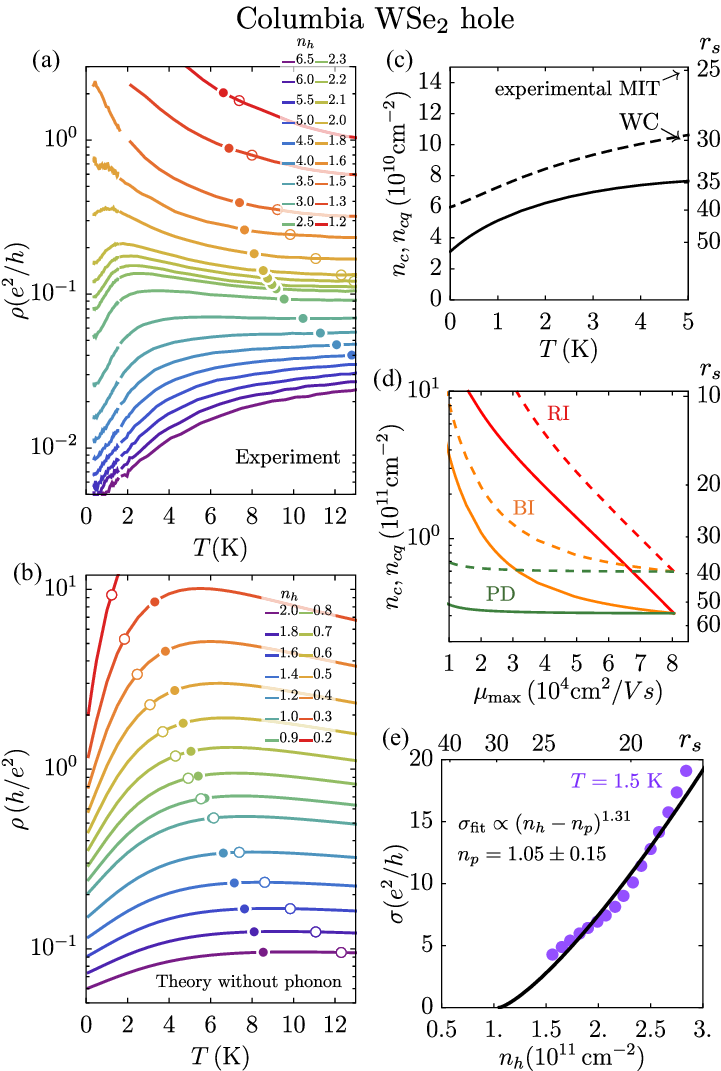}%mu_low_ne0.eps
    \caption{Analysis of the metal-insulator transition (MIT) at low densities of the Columbia sample. (a) shows the experimental temperature dependent resistivity data. (b) shows the theoretical result of impurity scattering , where the impurity parameters are taken from Fig.~\ref{fig:muT_n} (c). The inserts show the values of $n_h$ in units of $10^{11}$ cm$^{-2}$, with each density represented by a different color. The solid disks and empty circles mark $\rho(T=T_{\mathrm{BG}})$ and $\rho(T=E_F)$ respectively. (c) Temperature-dependent effective MIT critical densities $n_c(T)$ (solid) and $n_{cq}(T)$ (dashed). The right vertical axis labels the corresponding value of $r_s$. (d) $n_c$ and $n_{cq}$ vs maximum mobility at $T=0$. Different colors indicate the maximum mobility is tuned by adding either remote impurities (red), background impurities (orange), or atomic point defects (green) respectively. (e) Percolation fit of the low-$T$ conductivity data at low density.}
    \label{fig:rho_T}
\end{figure}

Another feature we observe in Fig.~\ref{fig:muT_n} is that, at large hole densities $n_h \gtrsim 2\times 10^{11}$ cm$^{-2}$, mobility decreases as the temperature increases.
This same feature leads to the linear-in-$T$ resistivity as we show in Fig.~\ref{fig:rho_T} (a-b).
%This is due to the linear $E_F \gg T$ $E_F \ll T$ and the system becomes a classical gas with a thermal velocity larger than the Fermi velocity, which leads to
On the other hand, from Fig.~\ref{fig:muT_n} at small densities $n_h \lesssim 2\times 10^{11}$ cm$^{-2}$, we see that $\mu$ increases monotonically as a function of temperature, which is related to the insulating behavior $d\rho/dT < 0$ depicted in Figs.~\ref{fig:rho_T} (a-b).
(However, it is important to note that data at low temperatures below 2 K for densities less than $1.5 \times 10^{11}$ cm$^{-2}$ is not available due to significant noise. 
This limitation may arise from problems such as contact resistance and experimental issues such as the stray capacitance of cryostat wiring and the decoupling of the electron temperature from that of the immersion cryogen, which compromise the reliability of measurements in this low-temperature, low-density regime.)
%raising the question of whether the resistivity decreases at very small temperatures.

\begin{figure}[t]
    \centering
    \includegraphics[width = \linewidth]{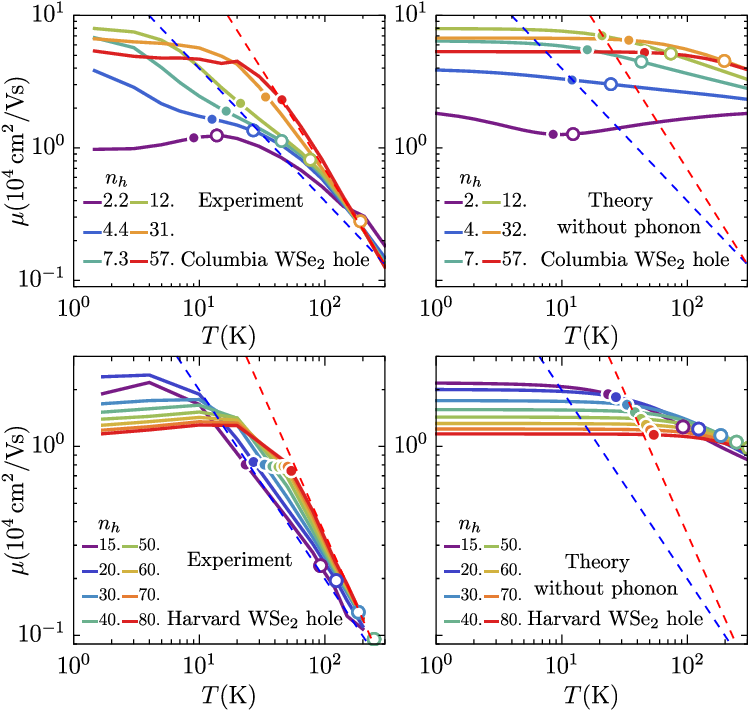}%mu_low_ne0.eps
    \caption{Mobility as a function of temperature at various densities. The solid disks and empty circles mark $\mu(T=T_{\mathrm{BG}})$ and $\mu(T=E_F)$ respectively. The different colors represent different densities $n_h$ in units of $10^{11}$ cm$^{-2}$. The dashed curves in blue and red serve as a guide to the eyes for the slopes $T^{-1}$ and $T^{-1.5}$ respectively. }
    \label{fig:mu_T}
\end{figure}

The observed temperature-dependent mobility and resistivity can be explained by the Boltzmann transport theory with screened charged impurity scattering.
At very low temperatures, the momentum relaxation of electrons, driven by their elastic scattering from impurities and structural defects in disorder metals, dominates the transport properties ~\cite{Ando_review:1982,Altshuler_Aronov:1985}.
At $T=0$, only scattering processes with momentum $q\leq 2k_F$ contribute to resistivity.
However, at finite temperatures, scattering with momentum greater than $2k_F$ becomes possible, where the non-analyticity of the 2D polarizability function at momemtum $q\geq 2k_F$ gives rise to real-space Friedel oscillations~\cite{Hwang_low_temp:2015}.
These Friedel oscillations, in turn, enhance backscattering through coherent interference of electrons scattered from these oscillations, thereby contributing to the observed linear-in-$T$ resistivity at low temperatures $t = T/E_F \ll 1$~\cite{Hwang_low_temp:2015}
\begin{align}\label{eq:rho_small_t}
    \rho(t \ll 1) \approx \rho_0 \qty(1 + \frac{2 s}{1+s}t ), 
\end{align}
where $\rho_0$ is the resistivity at $T=0$, and the screening strength is characterized by a dimensionless parameter $s=q_{TF}/2k_F$ with the Thomas-Fermi screening wave vector defined as $q_{TF} = g/a_B$.
Since we are interested in the temperature dependence of resistivity near the 2D MIT (the low density regime) where $k_F = \sqrt{4\pi n/g}$ is rather small compared to $q_{TF}$, we focus on the limit of large screening where $s \gg 1$.
%It should be emphasized that the interference effects produce a growth of resistivity as the temperature decreases.
On the other hand, at high temperatures $t \gg 1$, the system becomes a classical gas where the Thomas-Fermi screening is replaced by the Debye-H\"{u}ckel screening $q_{DH} = q_{TF}/t$~\cite{Fetter:1974,Ando_review:1982}.
To the leading order in $t \gg 1$, the Coulomb potential is unscreened at high temperatures with a Fourier component $U_0(q) = 2\pi e^2/q$, where the typical momentum scattering scale is determined by $\hbar^2 q^2/2m \sim T$ or $q \sim t^{1/2} k_F$.
Using Fermi's golden rule, the corresponding scattering rate $\tau^{-1} \propto U_0(q)^2 \propto t^{-1}$ results in a resistivity that decreases as the temperature rises~\cite{Hwang_low_temp:2015}
\begin{align}\label{eq:rho_large_t}
    \rho(t \gg 1) \approx \rho_0 s^2 t^{-1},
\end{align}
where an extra factor of $s^2$, in front of the $T=0$ resistivity $\rho_0$, takes care the fact that the charge is essentially unscreened at high temperatures $T\gg E_F$. 
In the calculations, we take into account the electron-electron interactions solely through screening, which is equivalent to the density-density (Hartree) interaction between the scattered electron and the Friedel oscillations that results into the linear-in-$T$ resistivity [cf. Eq.~(2.12) in Ref.~\cite{Zala_Aleiner:2001}].
Our theory of transport limited by scattering from screened disorder is motivated physically since the regularization of the long-range Coulomb potential is essential for sensible results. 
The theory may be construed as a mean field transport theory since the screened disorder is calculated at a mean field level using the RPA screening.  
%The exchange interaction contributes to the linear dependence of resistivity on temperature with a negative sign, as discussed in the Fock term Eq.~(2.13) in Ref.~\cite{Zala_Aleiner:2001}. 
%This effect is considered minor, at least in the metallic phase of the experimental data, given the consistently positive slope whose magnitude increases as the density decreases, leading us to exclude the exchange interaction from our calculations.
%We emphasize that this linearly $T$-dependent metallic resistivity in our theory (and presumably as observed experimentally in Ref.~\cite{CDean_WSe2:2023} does not arise from any $T = 0$ quantum criticality and does not imply any non-Fermi-liquid behavior. 
%It is simply a nonanalytic anomalous finite-$T$ property of the 2D Fermi surface, where the temperature dependence deviates qualitatively from that given by the Sommerfeld expansion arising from the $2k_F$ Kohn anomaly~\cite{Kohn:1959}. 
As mentioned before, any temperature dependence caused by phonon scattering is irrelevant within the temperature range of interest $T\lesssim 10$ K. 
This is due to the weak electron-phonon coupling in TMD materials and the negligible phonon-induced temperature variation at low temperatures, which scales as $(T/T_{BG})^4$ at temperatures well below the Bloch-Gr\"{u}neisen temperature $T_{BG} = 2\hbar v_s k_F$~\cite{Hwang_low_temp:2015,Lavasani:2019,Poduval_Kondo:2022}, where $v_s = 3.3 \times 10^5$ cm/s is the speed of sound in WSe$_2$~\cite{Jin_Zhenghe:2014}.
On the other hand, in the high-temperature range of 50--300 K, mobility decreases due to electron-phonon scattering, following a power law relationship $\mu \propto T^{-\gamma}$, where $\gamma$ ranges from 1 to 1.5.
We extrapolate the $T^{-1}$ contribution to the mobility coming from acoustic phonon scattering as shown by the red dashed lines shown in Fig.~\ref{fig:mu_T}, which is negligible for $T<10$ K by an order of magnitude compared to both our theoretical results and the experimental results of Refs.~\cite{CDean_WSe2:2023,PKim:2023}.
The blue dashed lines shown in Fig.~\ref{fig:mu_T} represent the $T^{-1.5}$ contribution coming from the optical phonon scattering, which also plays a minor role in the low-$T$ range.
%The typical temperature scale above which the phonon scattering becomes important roughly matches the value of $T_{\mathrm{BG}}$, which are labeled by solid disks in Fig.~\ref{fig:mu_T}. 
Since the competition between impurity and electron-phonon scattering is determined by two energy scales $E_F$ and $T_{BG}$, we mark the resistivity and mobility at $T=E_F$ and $T=T_{\mathrm{BG}}$ as empty circles and solid disks in Figs.~\ref{fig:rho_T} and \ref{fig:mu_T}, respectively.
We find that the typical temperature scale above which the phonon scattering becomes important roughly matches the value of $T_{\mathrm{BG}}$.
More detailed discussion of phonon scattering can be found in Sec.~\ref{sec:phonon}, but our focus in the current work is on the low-$T$ disorder scattering and Wigner crystallization.

Now, we discuss our theoretical results of the 2D MIT and the comparison with the experimental data of Columbia sample, as shown in Figs.~\ref{fig:rho_T} (c-e).
The MIT, fundamentally a zero-temperature phenomenon, is characterized by a critical density $n_c$ above which the system exhibits finite resistivity at $T=0$, and below which resistivity diverges exponentially as $T$ approaches zero, indicative of an insulating state~\cite{Shashkin_Kravchenko:2019}.
Ref.~\cite{CDean_WSe2:2023} suggests that the observed 2D MIT in monolayer WSe$2$ could be attributed to Wigner crystal formation, given a large value of $r_s \approx 25$ at the experimental critical density $n_c \approx 1.5 \times 10^{11}$ cm$^{-2}$.
However, we explore an alternative hypothesis where disorder, particularly from charged impurities, drives the MIT. 
This includes scenarios such as Anderson localization, where coherent electron scattering by random Coulomb impurities leads to localization, and classical percolation transition at finite temperatures, where electrons navigate through a landscape sculpted by long-range Coulomb potentials.
It is important to emphasize that a pristine WC is a perfect conductor, and any insulating transport in a WC must necessarily arise from impurities and disorder pinning the crystal, thus making transport in the WC regime inherently a disorder-dominated phenomenon.

Focusing first on Anderson localization, we use the Anderson-Ioffe-Regel (AIR) criterion~\cite{anderson1958,ioffe1960} to estimate $n_c$ by equating the mean free path $l = v_F \tau$ to the electron wavelength $\sim k_F^{-1}$ such that $k_F l = (2/g)(h n_h \mu/e) = 1$, or equivalently $\rho = (2/g) (h/e^2)$.
This criterion, when considered at finite temperatures, reveals a temperature-dependent effective critical density $n_c$, especially if the resistivity exhibits a strong temperature dependence. 
Our analysis shows that as temperature increases, the effective $n_c$ also increases, which is reflected in the resistivity trend observed in Fig.~\ref{fig:rho_T} (c).
Furthermore, we also estimate the effective critical density $n_{cq}(T)$ defined through $k_{F} l_q = (2/g)(h n_h \tau_{q}/m) = 1$, where $\tau_{q}$ is the quantum (single particle) scattering rate.
%In Fig.~\ref{fig:rho_T} (c) we show the effective $n_c$ and $n_{cq}$ as functions of $T$ which are roughly consistent with the data reported in Ref.~\cite{CDean_WSe2:2023}.
Our result of $n_{cq} = 6.0 \times 10^{10}$ cm$^{-2}$ calculated at $T=0$ is consistent with the value reported in Ref.~\cite{CDean_WSe2:2023} using $\tau_q\approx 1$ ps (or $\Gamma_q = \hbar/2\tau_q \approx 4$ K) experimentally measured from Shubnikov-de Haas oscillations.
Note that in general $\tau_q$ is less than or equal to $\tau$, defining the transport relaxation time (and $n_c$), and their difference could be large in 2D systems with strong Coulomb disorder since $\tau_q$ is limited by small angle scattering which does not affect $\tau$~\cite{Stern:1985}.
%\YH{To fit the experimental mobility data at low temperatures, we also find the in-plane charged impurity concentration should be smaller than $10^{9}$ cm$^{-2}$, in reasonable agreement with the charged impurity density measured by STM in crystals grown under similar conditions~\cite{liu2023twostep}, where they reported $n_i < 3.0\times 10^9$ cm$^{-2}$ for low-$T$ mobility exceeding $4.4 \times 10^{4}$ cm$^{2}$/Vs.}
To test the hypothesis that the origin of the 2D MIT is due to Anderson localization, we theoretically predict the AIR critical densities $n_c$ and $n_{cq}$ vs the maximum mobility $\mu_{\max}$ (peak mobility reached by tuning the carrier density) for different impurity scenarios, as shown in Fig.~\ref{fig:rho_T} (d).
Different colors indicate that the maximum mobility is adjusted by adding remote impurities (red), background impurities (orange), or atomic point defects (green), while keeping the other impurity concentrations fixed.
In all cases, $n_c$ and $n_{cq}$ increase as $\mu_{\max}$ decreases.
The addition of RIs leads to an exponential increase in $n_c$ with decreasing $\mu_{\max}$ (i.e., $n_c \simeq n_{r0} e^{-\mu_{\max}/\mu_{r0}}$ with $\mu_{r0} \simeq 2\times 10^{4}$ cm$^{2}$/Vs and $n_{r0} \simeq 1.6 \times 10^{12}$ cm$^{-2}$), whereas BIs result in a power-law relationship between $n_c$ and $\mu_{\max}$ (i.e., $n_c\propto \mu_{\max}^{-\alpha}$ with $\alpha$ changes from 0.5 to 2 as $\mu_{\max}$ decreases from 8$\times$ to 1$\times 10^4$ cm$^{2}$/Vs). 
Conversely, adding PDs only marginally affects $n_c$ (i.e., $n_c$ increases only by $\lesssim 20\%$ while $\mu_{\max}$ decreased by a factor of 10), suggesting that PDs play a minor role in MIT compared to other types of impurities. 
The fact that PDs play a relatively minor role in 2D TMD localization properties is a significant finding of our work.

Next, we discuss the scenario of MIT as a classical percolation transition. 
In the presence of long-range Coulomb disorder, electrons at low densities are separated into inhomogeneous metallic puddles by insulating barriers formed by Coulomb potential fluctuations~\cite{Thouless:1971,Shklovskii:1972,Kirkpatrick:1973,Shklovskii:1975,Ando_review:1982,shklovskii1984,DasSarma:2005579,Shklovskii:2007,Shaffique:2007,Manfra:2007,Tracy:2009,DasSarma:2005,Qiuzi:2013,Tracy:2014,Huang:2021a,Huang:2021,Huang:2022,Huang:2023}.
The ratio of metallic to insulating regions decreases as the MIT is approached from the metallic side, with the MIT signifying the transition to a regime in which insulating regions percolate and metallic regions become isolated puddles (whereas the situation is reversed in the conducting metallic regime-- that the MIT is characterized by the percolation transition where exactly one percolating path spans the whole device leading to metallic conductivity).
The percolation threshold of this MIT can be approximately estimated using $n_p \approx 0.1 \sqrt{n_r}/d \approx 1.6 \times 10^{11}$ cm$^{-2}$ where we use the remote impurity concentration $n_r = 3.0\times 10^{12}$ cm$^{-2}$ and $d=10$ nm obtained from the best-fit of the low-$T$ mobility data shown in Fig.~\ref{fig:muT_n} (c).
Here, only RI plays a role in the percolation transition and the BI is simply ignored; this situation is applicable only when RI is the dominant scattering mechanism at low carrier densities.
This expression of $n_p$ can be theoretically understood~\cite{Efros:19881019} as the typical density variation induced by random remote impurities within a square of $d \times d$, where the distance of remote impurities $d$ plays the role of the screening length.
The typical number fluctuation of random remote impurities is $\sqrt{n_r d^2}$ according to the Poisson distribution, so the corresponding density fluctuation is $\sqrt{n_r}/d$, which is the same as the above expression of $n_p$ up to a numerical factor. (This numerical factor is obtained by computational simulations in Ref.~\cite{Efros:1993}.)
Assuming that the scaling of the 2D percolation conductivity follows $\sigma \propto (n_h - n_p)^{1.31}$~\cite{Kirkpatrick:1973,Isichenko:1992,DasSarma:2005579,Tracy:2009,DasSarma:2005,Manfra:2007,Qiuzi:2013,Tracy:2014}, we obtain the percolation threshold $n_p=(1.05 \pm 0.15) \times 10^{11}$ cm$^{-2}$ by fitting the low-$T$ data at low densities $n_h < 3 \times 10^{11}$ cm$^{-2}$.
See Fig.~\ref{fig:rho_T} (e).
We find both the AIR $n_c$ and the percolation threshold $n_p$ have high values of $r_s$ close to the theoretical WC transition predicted by the QMC calculation in the absence of disorder, where $r_s \approx 30$~\cite{Drummond:2009}, suggesting that the observed 2D MIT behavior likely results from the complex interplay between disorder effects and interaction-driven WC physics, since at these low carrier densities both disorder and correlation are important.
\begin{figure*}[t]
    \centering
    \includegraphics[width = \linewidth]{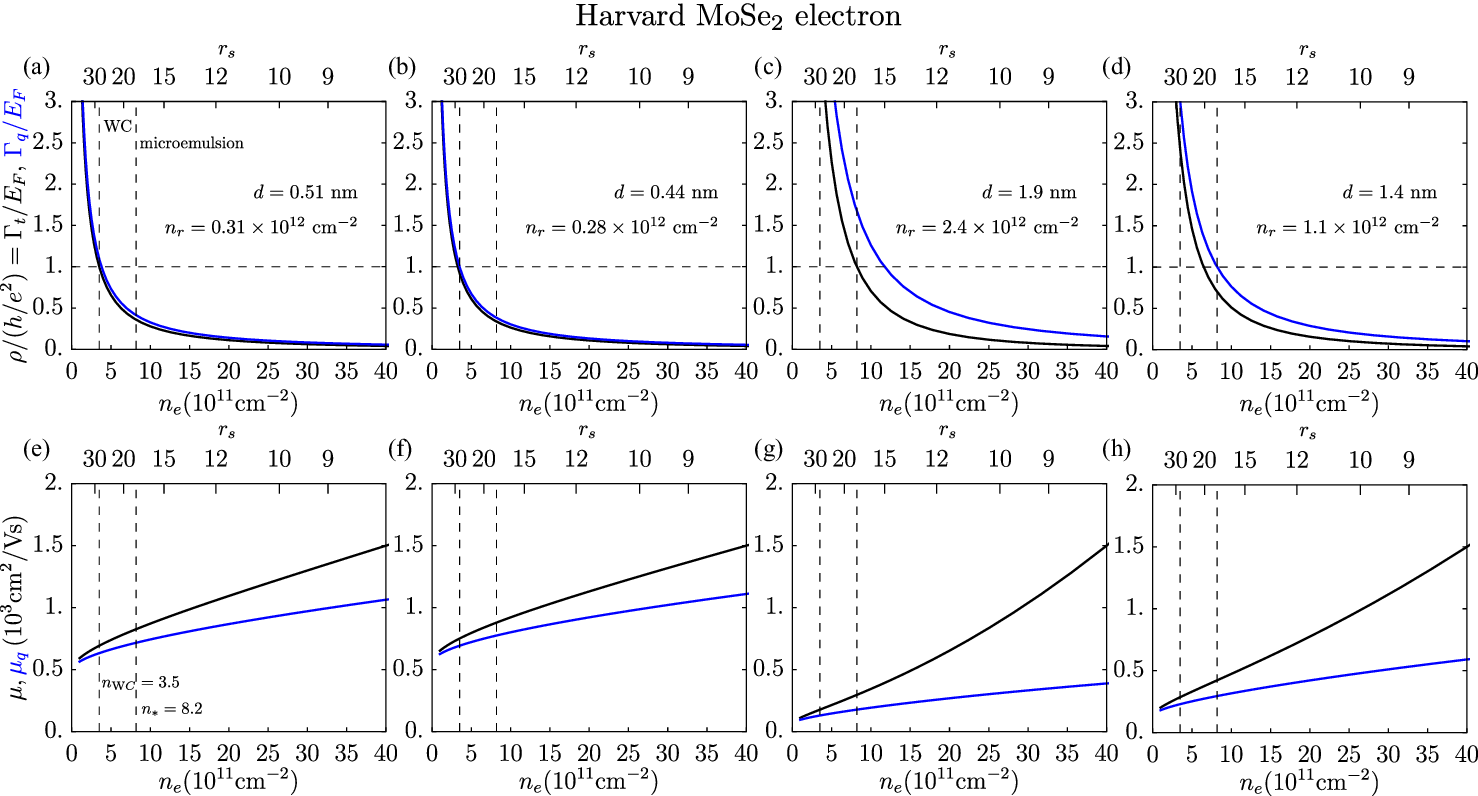}%mu_low_ne0.eps
    \caption{Theoretical analysis of zero-temperature resistivity and mobility as a function of density for Harvard experiments on monolayer MoSe$_2$. (a), (b), (c), and (d): Dimensionless resistivity and scattering rates as a function of density. The black curves are $\rho / (e^2/h) = \Gamma_t/E_F$, while the blue curves are $\Gamma_q/E_F$, where $\Gamma_t = \hbar / 2\tau$ and $\Gamma_q = \hbar / 2\tau_q$ are the transport and the single-particle scattering rate, respectively. The vertical dashed lines mark the Wigner crystal transition density $n_{\mathrm{WC}} = 3.5 \times 10^{11}$ cm$^{-2}$ and the microemulsion to Fermi-Liquid transition density $n_{*} = 8.2 \times 10^{11}$ cm$^{-2}$. The horizontal dashed lines mark the AIR condition $\rho / (e^2/h) = \Gamma_t/E_F = 1$ and $\Gamma_q/E_F = 1$. (e), (f), (g), and (h) show the corresponding mobility (black) and quantum mobility (blue) as a function of density. In other words, (a) and (e) show $\rho / (e^2/h) = \Gamma_t/E_F = 1$ at $n_{\mathrm{WC}}$, (b) and (f)  show $\Gamma_q/E_F = 1$ at $n_{\mathrm{WC}}$, (c) and (g) show $\rho / (e^2/h) = \Gamma_t/E_F = 1$ at $n_{*}$, (d) and (h) show $\Gamma_q/E_F = 1$ at $n_{*}$. The obtained impurity parameters $n_r$ and $d$ are shown directly in the figures.}
    \label{fig:optimal_d}
\end{figure*}

Finally, we comment on the MoSe$_2$ electron system studied by the second Harvard group~\cite{sung2023observation}.
Reference~\cite{sung2023observation} reports a quantum crystal-to-liquid transition, where the signature of a mixed state between the Wigner crystal and Fermi liquid, the microemulsion phase, is observed through the anomalies in exciton reflectance, spin susceptibility, and umklapp scattering.
In the absence of disorder, this variation in local electron density can arise from the Coulomb-frustrated phase separation~\cite{Spivak_Kivelson:2004,Jamei:2005,Spivak_Kivelson:2006,Li_Shiqi:2019}, leading to the microemulsion phase.
However, when the electron density is small and close to the MIT critical density $n_c$, there are not enough electrons to screen the long-range Coulomb disorder, and the electrons break into inhomogeneous metallic puddles embedded in the insulating potential barriers~\cite{Shklovskii:1972,Ando_review:1982,shklovskii1984,Shklovskii:2007,Shaffique:2007}, potentially smearing the transition into what experimentally appears as a microemulsion mixed state~\cite{Joy:2023}.
Thus, microemulsion at low densities is practically the same as Coulomb disorder induced puddles and percolation scenario, and cannot be distinguished easily.
In Ref.~\cite{sung2023observation}, instead of transport measurement that probes the macroscopic sample globally, cryogenic reflectance and magneto-optical spectroscopy are used to focus on a local spot of size $\sim 760$ nm, determined by the laser diameter of the diffraction limit.
This leads to several important questions that we should address in the following discussion. 
What would the corresponding transport data look like? 
What impact does disorder have on the MoSe$_2$ electron system? 
We emphasize that even in the cleanest samples it is not possible to completely eliminate the role of Coulomb disorder, which tends to produce long-range variation in the local electron density, causes the 2D electrons to break up into itinerant and localized regions, and leaves the WC phase with only finite-range order~\cite{Falson2022}. 
Figure~\ref{fig:optimal_d} presents the theoretical result of zero-temperature resistivity and mobility as a function of density for monolayer MoSe$_2$ electron system.
We model the disorder as a delta layer of remote impurities of concentration $n_r$ located at a distance $d$ away from the TMD monolayer. 
Since there are no experimental transport data in Ref.~\cite{sung2023observation} for comparison, to obtain an estimate of $n_r$ and $d$, we make two assumptions in the calculation. 
First, we assume the mobility $\mu = 1.5 \times 10^3$ cm$^{2}$/Vs at high electron density $n_e = 4.0\times 10^{12}$ cm$^{-2}$. 
This is based on the knowledge that the MoSe$_2$ sample used in Ref.~\cite{sung2023observation} is synthesized by the CVT method, which generally has a peak mobility of around 1,000 cm$^{2}$/Vs at low temperatures~\cite{PKim:2023}.
Second, we assume that the AIR condition $\rho / (e^2/h) = \Gamma_t/E_F = 1$ or $\Gamma_q/E_F = 1$ is met at a density somewhere within the range of the experimental reported densities of WC formation $n_{\mathrm{WC}} = 3.5 \times 10^{11}$ cm$^{-2}$ and the microemulsion to Fermi-liquid transition $n_{*} = 8.2 \times 10^{11}$ cm$^{-2}$.
Here, $\Gamma_t = \hbar / 2\tau$ and $\Gamma_q = \hbar / 2\tau_q$ are the transport and the single-particle scattering rates, respectively.
Using these two constraints, we obtain the estimate of $n_r\sim 10^{11}$--$10^{12}$ cm$^{-2}$ and $d\sim $0.5--2 nm, as shown in the inserts of Figs.~\ref{fig:optimal_d} (a-d).
Our estimate is consistent with recently reported STM images of disordered pinned Wigner crystal of short-range crystalline order and the random charged defects in bilayer MoSe$_2$, where the average distance between adjacent impurities is $\sim20$ nm, corresponding to a charged impurity density $\sim 3 \times 10^{11}$ cm$^{-2}$~\cite{xiang2024quantum}.
In the presence of these Coulomb impurities, the disorder smearing of the transition which may appear to be the intermediate microemulsion phase should also cover a range of $\Delta n_e \sim n_r$, which is consistent with the experimental range $n_{*}-n_{\mathrm{WC}} =4.7 \times 10^{11}$ cm$^{-2}$.
We note that the average distance between impurities $n_r^{-1/2} \sim $10--30 nm is smaller than the reported correlation length $l_{\mathrm{corr}}\sim 3 a_{\mathrm{WC}}\sim 70$ nm for the short-range crystalline order WC, extracted from the lineshape of the umklapp resonance~\cite{Smolenski:2021}, where $a_{\mathrm{WC}}\sim n_e^{-1/2}$ is the WC lattice constant.
This indicates that the laser spot reported in Ref.~\cite{sung2023observation} probably represents a local region in the sample which is ``cleaner'' than the average.
(The possibility that this factor of 2 difference is not significant, arising simply from the unknown system parameters, cannot be ruled out here.)
Considering that these impurity parameters are reasonable and agree with the previous estimate in similar samples~\cite{Ahn_temperature_mTMD:2022,Ahn_MIT_mTMD:2022}, and the impurity concentration is close to the electron densities in the range of interest ($n_{*}$--$n_{\mathrm{WC}}$), disorder should strongly affect the transport measurement results.
On the other hand, we notice that the melting temperature of WC $\sim$10 K reported in Refs.~\cite{Smolenski:2021,sung2023observation} is much higher than the value $\sim 0.8$ K predicted by a mean-field theory in the absence of disorder~\cite{Hwang_WC:2001} (see the detailed discussion in Sec.~\ref{sec:WC_melting}).
This phenomenon can be explained by the fact that the disordered crystal at low carrier density often transforms into an insulating glassy state that non-perturbatively incorporates both Anderson localization and Wigner crystallization physics, which strongly enhances the stability of the insulating phase and increases its melting temperature, as the glassy phase is much more thermally stable in general~\cite{DinhDuy_Vu:2022}.

Our theoretical findings depicted in Figs.~\ref{fig:muT_n}, \ref{fig:rho_T}, and~\ref{fig:mu_T} manifest a qualitative and semi-quantitative agreement with the experimental data of WSe$_2$ hole system presented in Refs.~\cite{CDean_WSe2:2023,PKim:2023}. 
Moreover, we discuss the role of disorder and predict the transport resistivity and mobility for the MoSe$_2$ electron system reported in Ref.~\cite{sung2023observation}.
Specifically, our theory successfully captures three key characteristics seen in the experimental data: (1) a linear-in-$T$ metallic resistivity at low temperatures; (2) an enhanced slope $d\rho/dT$ as the carrier density approaches the MIT critical density from the metallic side; and (3) an increase in the metallic resistivity by a factor of approximately 2-6 with a modest temperature increment of about 10 K.
We should mention that the theoretical density range plotted in Figs.~\ref{fig:rho_T} is smaller than the experimental density range.
The reasons are twofold. 
First, the theoretical AIR MIT happens at a smaller critical density compared to the experimental MIT because of the failure of the Boltzmann theory with linear screening at low densities.
As a result, we need to systematically shift the range of carrier density to capture the regime near the theoretical AIR MIT.
Second, given the inhomogeneous electron puddle picture near the critical density, the whole system is divided into metallic and insulating regions, and the actual Fermi surface made by itinerant electrons is smaller than the one calculated using apparent carrier density (see a two-species model for transport in Ref.~\cite{Li_Qiuzi:2013}).
Therefore, near the MIT, the effective carrier density used in the Boltzmann transport calculation should be smaller than the experimental density due to carrier trapping in puddles.
Finally, we note that due to its atomic thinness, monolayer TMD does not have interfacial roughness, a scattering mechanism that typically restricts mobility at high densities in materials like GaAs and Si quantum wells~\cite{CDean_WSe2:2023,Huang:2022a,huang2023understanding}.
This suggests that the mobility of monolayer TMDs, eventually limited by unintentional charged impurities and atomic point defects, holds significant potential for enhancement through improved synthesis techniques and the careful assembly of heterostructure devices so as to suppress the impurity content.

\section{Melting of Wigner crystals}
\label{sec:WC_melting}
One of the most well known examples of a quantum phase transition is the liquid-solid transition for electrons at $T=0$, as predicted by Wigner 90 years ago~\cite{Wigner:1934}, also known as the Wigner cyrstal (WC) transition, which, in its classical version of thermal melting, was first demonstrated in a low-density 2D electron system realized on the Helium surface~\cite{Grimes:1979}.
[This system had such an extreme low electron density that the resulting electron (Wigner) crystal was classical ($T>E_F$), whereas the original work by Wigner as well as the current interest in WC of 2D semiconductor and TMD layers is in the $T=0$ quantum WC ($T<E_F$).]
Since then, significant efforts have been made to explore the WC transition in semiconductor-based  platforms.
At high densities, electrons arrange themselves into a metallic liquid to minimize the quantum kinetic energy, whereas at lower densities, electrons are anticipated to solidify into an insulating WC when the Coulomb interaction energy significantly outweighs the kinetic energy.
In this section, we focus on the 2D WC and briefly discuss its melting.
(One-dimensional WCs consisting of a few electrons have been imaged in a carbon nanotube experimentally~\cite{Shapir:2019}, and have been theoretically studied in Ref.~\cite{DinhDuy_Vu:2020}.)
\begin{figure}[t]
    \centering
    \includegraphics[width = \linewidth]{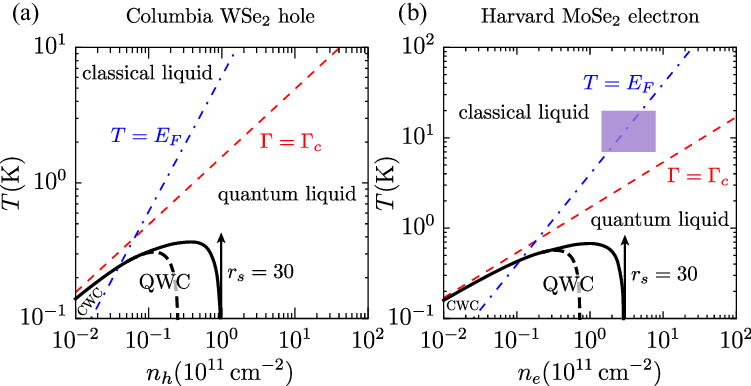}%mu_low_ne0.eps
    \caption{The density-temperature mean-field phase diagram of the disorder-free pristine electron solid-to-liquid transition, calculated using material parameters for (a) Columbia WSe$_2$ hole system and (b) MoSe$_2$ electron system. The shaded area shows the region of experimentally reported melted temperatures~\cite{Smolenski:2021}. $\Gamma_c = 120$~\cite{Gann:1979}.}
    \label{fig:WC_phase}
\end{figure}
\begin{table*}[t]
\caption{Melting temperatures of WCs in different materials, where e (h) represents electron (hole). (``BLG'' represents bilayer graphene.) $\kappa$ and $m$ are the dielectric constant and the effective mass, respectively. The columns of $n$ and $r_s$ label the range where WCs are reported in the corresponding materials. $T_e$ is the lowest electronic temperature achieved in experiments. ``exp. $T_m$'' represents the experimental melting temperature, while ``MF $T_m$'' represents the highest theoretical melting temperature $\approx (e^2/\kappa a_B)/(4\Gamma_c r_c)$ predicted by mean field (MF) theory, where $\Gamma_c = 120$~\cite{Gann:1979} and $r_c = 30$~\cite{Drummond:2009}.}
\begin{ruledtabular}
\begin{tabular}{c | c | c | c | c | c | c | c} 
 material & $\kappa$ & $m/m_0$ & $n$  & $r_s$ & $T_e$ & exp. $T_m$ & MF $T_m$ \\
%heading
\hline \\ [-3.2ex] 
%\hline \\ Shayegan:2020,Pfeiffer:2018
WSe$_2$ h~\cite{CDean_WSe2:2023} &5 & 0.45 & 1.2--1.5$\times 10^{11}$ cm$^{-2}$ & 25--28 & 0.5 K & not available & 0.4 K \\ 
MoSe$_2$ e~\cite{Smolenski:2021,sung2023observation} &4.6 & 0.7 & $\lesssim$5.0$\times 10^{11}$ cm$^{-2}$ & $\gtrsim$23 & 80 mK & 10 K & 0.8 K \\ 
AlAs e~\cite{Hossain:2020,Hossain:2022} & 10.9 & 0.46 & 1.0--1.8$\times 10^{10}$ cm$^{-2}$ & 38--50 & 300 mK & 1 K & 90 mK \\ 
ZnO e~\cite{Falson2022} & 8.5 & 0.3 & 1.0--1.6$\times 10^{10}$ cm$^{-2}$ & 30--38 & 10 mK & 50 mK & 90 mK \\
GaAs h~\cite{Shayegan:1999} & 12.9 & 0.5~\cite{Pfeiffer:2007} & 0.48--3.72$\times 10^{10}$ cm$^{-2}$ & 21--60 & 50 mK & 145 mK & 70 mK \\
GaAs h~\cite{Shayegan:1991,Pfeiffer:2018,Shayegan:2020} & 12.9 & 0.5~\cite{Pfeiffer:2007} & 2.0--7.9$\times 10^{10}$ cm$^{-2}$ & 15--30 & 10--40 mK & 35--400 mK & 70 mK \\
GaAs e~\cite{Goldman:1990,Pfeiffer:2006,Shayegan:2019} & 12.9 & 0.067 & 1.2--8.1$\times 10^{10}$ cm$^{-2}$ & 2--5 & 10--40 mK & 50--900 mK & 10 mK \\ 
BLG e~\cite{tsui2023direct} & 3~\cite{bbg_dielectric} & 0.041~\cite{Zou:2011} & 3.1--9.0$\times 10^{9}$ cm$^{-2}$ & 15--26 & 210 mK & 3 K & 100 mK 
\end{tabular}
\end{ruledtabular}
\vspace{-0.2 in}
\label{table:WC_melting}
\end{table*}

We first derive the density-temperature mean-field phase diagram of the pristine 2D WC transition shown in Fig.~\ref{fig:WC_phase}. 
At low temperatures $T<E_F$, the so-called quantum regime, the kinetic energy $\ev{K}$ is given by the zero point motion of the crystal $\sim E_F\propto n$; while in the classical regime at $T>E_F$, the dominant contribution to $\ev{K}$ comes from the thermal motion $\sim T \propto n^0$.
Meanwhile, the Coulomb interaction energy (the cohesive energy of the solid), expressed as $\ev{V} = (e^2/\kappa)\sqrt{\pi n}$, is directly proportional to $\sqrt{n}$, indicating that $\ev{V}$ always dominates over $\ev{K}$ at a sufficiently low density and temperature, causing the electrons to solidify into a WC.
On the other hand, if $\ev{K} > \ev{V}$, then WC melts into an electron liquid.
This leads to a WC dome in the temperature-density phase diagram. See Fig.~\ref{fig:WC_phase}.
At $T=0$, there is a quantum phase transition at $r_s = \ev{V}/E_F = r_c$ or $n=n_c = (\pi a_B^2 r_c^2)^{-1} \propto m^2/\kappa^2$, where $r_c \approx 30$ suggested by quantum Monte Carlo simulations~\cite{Drummond:2009}.
%For a material with larger effective mass and smaller dielectric constant, the critical density $n_c$ is larger so that it is easier to observe this quantum phase transition in the experimental density range. 
At finite temperatures, the classical liquid-solid transition occurs at $\Gamma = \ev{V}/T = \Gamma_c$ where $\Gamma_c \approx 120$ is found from molecular dynamics simulations~\cite{Gann:1979} and experimentally confirmed in liquid Helium~\cite{Grimes:1979}.
A simple way to interpolate between the classical and quantum liquid-solid phase transition boundaries is to replace the definition of $\Gamma$ by $\ev{V}/\ev{K}$, and smoothly change the value of $\Gamma_c$ to $r_c$ as $n$ approaches $n_c$~\cite{Hwang_WC:2001}.
Here, the average kinetic energy at a finite temperature is given by
\begin{align}
    \ev{K} = \frac{\sum_{\vb{k}} \epsilon(k) n_0(\epsilon)}{\sum_{\vb{k}} n_0(\epsilon)}
\end{align}
where $n_0(\epsilon) = [e^{(\epsilon - \mu)/T} +1]^{-1}$ is the Fermi-Dirac distribution.
The thick black solid lines in Fig.~\ref{fig:WC_phase} are examples of such interpolation results calculated for WSe$_2$ hole and MoSe$_2$ electron systems (the thick black dashed lines show the corresponding results for fixed $\Gamma_c = 120$).
Consequently, the highest melting temperature predicted by this mean-field theory is given by $T_m \approx (e^2/\kappa a_B)/(4\Gamma_c r_c) \propto m/\kappa^2$.
Therefore, it is more likely to observe WCs in the experimentally accessible temperature and density range for a material with a large effective mass and a small dielectric constant, so that both $T_m$ and $n_c$ are sufficiently large. 
As a result, TMDs with a large effective mass $m\sim 0.5 m_0$ and small dielectric constant in the hBN environment $\kappa \sim 5$ are generally much better platforms (than, e.g., GaAs electron system where $m \sim 0.067 m_0$ and $\kappa \sim 13$) for observing WCs~\cite{Smolenski:2021,sung2023observation}. 
Table~\ref{table:WC_melting} reviews the material platforms where WCs are reported experimentally.
Indeed, the monolayer MoSe$_2$ electron system with a critical density $\lesssim 5 \times 10^{11}$ cm$^{-2}$ and an experimental melting temperature $T_m \sim$10 K is by far the best material platform to observe WCs.

Next, we discuss the disorder effects on the thermal (i.e., temperature-induced at fixed low density) and quantum (i.e., density-induced at fixed low $T$) melting of WC.
The main effects are two-fold.
First, disorder can significantly increase the effective melting temperature because the disordered crystal often transitions into a localized glassy state that combines Anderson localization and Wigner crystallization physics, leading to an insulating state with enhanced stability~\cite{DinhDuy_Vu:2022}.
Second, in the presence of disorder, it becomes quite challenging to obtain a material that is clean enough to achieve such large $r_s$ (small $n_c$) values, so that the interaction-driven WC transition is not completely hindered by the single-particle localization.
We elaborate on our reasoning as follows.
From Table~\ref{table:WC_melting}, we find that the experimental melting temperature is generally much higher than the value predicted by the mean field theory for the pristine system.
This is because the unavoidable disorder present in the experiments tends to pin the WC, enhancing its stability.
%The experimental signal of pinned WCs is frequently linked to the non-linear dc current-voltage characteristics, and the melting point is determined as the temperature at which this nonlinear behavior disappears.
%The depinning threshold voltage of the WC, characterized by a sudden increase in current as the applied voltage exceeds the threshold, is proportional to the impurity density.
%However, this technique provides indirect evidence for WC formation, making it difficult to distinguish interaction-driven crystallization from localization by disorder.
Indeed, numerical studies show that the effective melting temperature increases as the disorder becomes stronger~\cite{DinhDuy_Vu:2022}.
We should emphasize that even in the absence of Coulomb interaction, there is a metal-insulator transition by lowering the electron density due to the single-particle Anderson localization.
For materials with a small critical density of the WC transition, the small number of electrons cannot sufficiently screen the Coulomb disorder potential, and the system becomes Anderson-localized before the WC transition.
This is why even fairly clean GaAs systems become highly insulating when $n\sim$8--9 $\times 10^9$ cm$^{-2}$ corresponding to $r_s \sim 15$ which is much lower than that necessary for WC~\cite{Simmons:1998,Shayegan:1999}.  
This emphasizes that associating low-density metal-insulator transitions automatically with WC, as is often done uncritically in the literature, is unfounded since the cleanest material (i.e., 2D n-GaAs with mobilities well in excess of $10^7$ cm$^2$/Vs) reflects a metal-insulator transition at an $r_s$ value far below the putative WC transition point.
Besides directly reducing the carrier density, an alternative to suppress the kinetic energy (to produce a WC) is to utilize the superlattice moir{\'e} potential, where the so-called ``generalized WC'' was imaged in moir{\'e} TMDs WSe$_2$/WS$_2$ heterostructures~\cite{Feng_Wang:2021}.
However, these electron phases with a crystalline structure differ from the WC phase that is anticipated to emerge naturally without any periodic potential, as they break a discrete rather than a continuous translational symmetry.
We emphasize that the reported ``WC'' in moir{\'e} TMD is basically a commensurate charge density wave since the lattice constant for this generalized WC is commensurate with the underlying moir{\'e} lattice whereas, by contrast, the WC in continuum TMD systems (which is the subject matter of the current work) has a lattice constant determined entirely by the electron density (with a lattice constant $\sim n^{-1/2}$ which is much larger than the ionic lattice size) with no connection to the underlying ionic lattice~\cite{haining_phase_moire:2020}. 
The connection between the continuum WC and lattice physics has been elucidated in Ref.~\cite{DinhDuy_Vu_dots:2020}.
Note that while a strong external magnetic field can also suppress kinetic energy to obtain WCs~\cite{Shayegan:1991,Pfeiffer:2018,Shayegan:2020,Goldman:1990,Pfeiffer:2006,Shayegan:2019,tsui2023direct}, in this paper we focus specifically on WCs at zero magnetic field.
We should emphasize that observations of WC at $r_s > 30$ without a magnetic field are quite rare due to the requirement for far more dilute electron densities where the disorder effect is even more prominent because of weaker screening.
This scenario is relevant to previous generations of TMDs grown by CVT methods, where the mobility is $\sim10^3$ cm$^{2}$/Vs with charged impurity density $\sim10^{11}-10^{12}$ cm$^{-2}$ close to the electron critical density $n_c$~\cite{Smolenski:2021,sung2023observation}.
In order for the WC transition to be no longer overwhelmed by the disorder-induced Anderson localization, it is necessary to enhance the sample quality and increase the mobility by reducing disorder.
The research groups at Columbia and Harvard are exploring this direction, demonstrating an order of magnitude improvement in the mobility of WSe$_2$ monolayers~\cite{CDean_WSe2:2023,PKim:2023}.
The quality of the sample is also reflected in the fact that the estimated critical density of the Anderson localization transition $\sim 6 \times 10^{10}$ cm$^{-2}$ is lower than the experimental critical density of MIT $\sim 1.5 \times 10^{11}$ cm$^{-2}$~\cite{CDean_WSe2:2023}.
Although the difference between these critical densities is a factor of $\sim2$, suggesting that disorder cannot be completely ignored, this new generation of TMD monolayers with record high mobility should be by far the best material platform for studying the WC physics, and most likely, a continuum WC has already been observed in clean 2D TMDs.

\section{2D polarizability and Friedel oscillations at finite temperatures}
\label{sec:review_polarizability}
\begin{figure}[t]
    \centering
    \includegraphics[width = \linewidth]{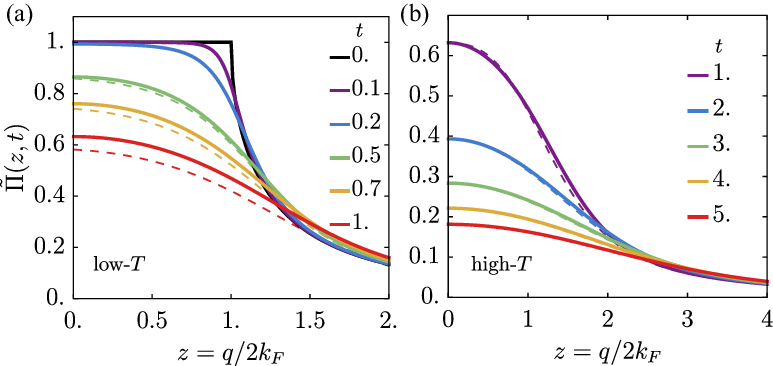}%mu_low_ne0.eps
    \caption{Dimensionless polarizability $\tilde{\pi}(z,t)$ as a function of $z = q/2k_F$ at (a) low temperatures $t\leq 1$ and (b) high temperatures $t\geq 1$. The solid curves are the complete numerical result calculated using Eq.~\eqref{eq:pi_zt}. The dashed curves are the analytical approximations using (a) Eq.~\eqref{eq:pi_zt_lowt} and (b) Eq.~\eqref{eq:pi_zt_hight_fullz}.}
    \label{fig:pi_t}
\end{figure}

We provide the temperature-dependent 2D polarizability and Friedel oscillations in this section.
Theoretical results on 2D polarizability as a function of temperature have been explored in Refs.~\cite{Stern:1967,Maldague:1978,Gold_Dolgopolov:1986,Ando_review:1982,Zala_Aleiner:2001,Hwang_finite_T:2003,Hwang:2004,DasSarma:2005579,Rossi:2011,Lavasani:2019}.
Despite this rich background, we obtain some analytical results, Eqs.~\eqref{eq:pi_zt_lowt} and \eqref{eq:pi_zt_hight_fullz} as approximations for the low- and high-temperature polarizability, which are good even at intermediate temperatures $T \sim E_F$, and should be useful in future theoretical works. (See Fig.~\ref{fig:pi_t}.)
As far as we know, Eqs.~\eqref{eq:pi_zt_lowt} and ~\eqref{eq:pi_zt_hight_fullz} have not been previously published, so we give a comprehensive derivation of these equations here.
%They are directly related to the experiments back to 1994–95 by Kravchenko and collaborators~\cite{Kravchenko:1994,Kravchenko_Si:1995,Kravchenko_QHE:1995,Kravchenko_comment:1999}, which created the modern subject of 2D MIT, serving as the temporal milestone separating the early days of 2D MIT~\cite{Ando_review:1982} from the present days of 2D MIT~\cite{Abrahams_Kravchenko_Sarachik:2001,Kravchenko_Sarachik:2004,Spivak_Kravchenko_Kivelson_Gao:2010}.
%The 2D MIT phenomenon was studied in the early days~\cite{Mott:1975,Adkins:1978,Ando_review:1982}, had high disorder (and low mobility) and consequently high $n_c$ (and $T_D$) leading to large $T_F \gg T_{\mathrm BG}$ (as well as large $T_D > T_{\mathrm BG}$) in the metallic phase ($n > n_c$) so that no metallicity could be observed except for phonon scattering effects for $T>T_{\mathrm BG}$. Thus, the amount of disorder in the sample leading to low or high $n_c$ (and $T_D$) is the key to the manifestation of a strong metallic temperature dependence in $\rho(T)$ for $n > n_c$.
%Transition to an insulator was discussed in Refs.~\cite{Hwang:2014a,Tracy:2009}.
%The temperature dependence of the polarization function is derived in Refs.~\cite{Maldague:1978,Gold_Dolgopolov:1986}.
The non-interacting polarizability at finite temperatures is defined as
\begin{align}\label{eq:polarization_def}
    \Pi(q,\omega; T, \mu) = g \int \frac{d^2 k}{(2\pi)^2} \frac{n_{k+q} - n_{k}}{\omega - \epsilon_{k+q} + \epsilon_{k}}.
\end{align}
where $g$ is the total quantum degeneracy of the 2D system.
$n_k$ is the Fermi-Dirac function given by
\begin{align}\label{eq:fermi-dirac}
    n_k = \frac{1}{e^{\epsilon_k - \mu} + 1} = \int \limits_0^{\infty} d\mu' \frac{\Theta(\mu' - \epsilon_k)}{4T \cosh^2[(\mu - \mu')/2T]},
\end{align}
where $\Theta(x)$ is the Heaviside-step function.
Substituting Eq.~\eqref{eq:fermi-dirac} into Eq.~\eqref{eq:polarization_def}, we arrive at~\cite{Maldague:1978}
\begin{align}
    \Pi(q,\omega; T, \mu) = \int  \limits_0^{\infty} d\mu' \frac{\Pi(q,\omega; T=0, \mu)}{4T \cosh^2[(\mu - \mu')/2T]},
\end{align}
The zero-temperature polarizability reads
\begin{align}
    \Pi(q,\omega; 0, \mu) = \int \frac{g d^2 k}{(2\pi)^2} \frac{\Theta(\mu' - \epsilon_{k + q}) - \Theta(\mu' - \epsilon_k)}{\omega - \epsilon_{k+q} + \epsilon_{k}}.
\end{align}
For parabolic energy dispersion $\epsilon_k = \hbar^2 k^2/2m$, the analytical expression of $\Pi(q,\omega; T=0, \mu)$ is derived in Ref.~\cite{Stern:1967}.
In the following, we focus on the static polarizability at $\omega = 0$, since this is what defines the screening of Coulomb disorder.
The static polarizability at $T = 0$ reads
\begin{align}\label{eq:pi_t0}
    \Pi(q,\mu) = g \frac{m}{2\pi \hbar^2} \qty[1 - \sqrt{1 - \frac{8 m \mu}{\hbar^2 q^2}} \Theta(q^2 - 8m \mu /\hbar^2)].
\end{align}
Introducing dimensionless notations ${\tilde \Pi} = \Pi/(g m/2\pi \hbar^2)$, $z = q/2k_F$, and $t = T/E_F$ where $E_F = \hbar^2 k_F^2 /2m$ is the Fermi energy defined at zero temperature, the $T=0$ dimensionless static polarizability Eq.~\eqref{eq:pi_t0} can be rewritten as 
\begin{align}\label{eq:tilde_pi_t0}
    {\tilde\Pi}(z) = 1 - \sqrt{1-z^{-2}} \Theta(z^2 - 1).
\end{align}
At finite temperatures, the polarizability reads
\begin{align}
    &{\tilde \Pi}(z,t) = \frac{1}{2}\qty(1 + \tanh{\frac{\mu}{2T}}) - \int \limits_0^{z^2/t} \frac{dx \sqrt{1 - t x / z^2} }{4 \cosh^2[(\mu/T - x)/2]}, \nonumber \\
    &= 1 - e^{-1/t} - \int \limits_0^{z^2/t} \frac{(\sqrt{t}/z) \sqrt{x} dx}{4 \cosh^2\{[\ln(e^{1/t} - 1) -z^2/t + x]/2\}}, \label{eq:pi_zt}
\end{align}
where we use the expression of the chemical potential $\mu$ 
\begin{align}
    \frac{\mu}{T} = \ln(e^{E_F/T} - 1).
\end{align}
At $z=0$, the integral in Eq.~\eqref{eq:pi_zt} vanishes so that the polarizability is given by
\begin{align}\label{eq:pi_z0}
    {\tilde \Pi}(z=0,t) = 1 - e^{-1/t}.
\end{align}
At low temperatures $t = T/E_F \ll 1$,
%we can extend the upper bound of the integral in Eq.~\eqref{eq:pi_zt} and arrive at
we can expand the integrand of Eq.~\eqref{eq:pi_zt} using 
\begin{align}\label{eq:cosh_expansion_low_T}
    \frac{1}{4\cosh^2(x/2)} = -\sum \limits_{n=1}^{\infty} n(- e^{-x})^n,\, x>0.
\end{align}
so that the polarizability becomes
\begin{align}
    {\tilde \Pi}(z,t\ll1) = 1 + \frac{\sqrt{\pi t}}{2 z} \sum_{n=1}^{\infty} \frac{e^{nz^2/t} \mathrm{erf}\qty(\sqrt{nz^2/t})}{n^{1/2} (1 - e^{1/t})^n}.
\end{align}
where $\mathrm{erf}(x)$ is the error function with an asymptotic expansion at large $x$
\begin{align}
    \mathrm{erf}(x) = 1 - \frac{e^{-x^2}}{\sqrt{\pi} x} \sum_{m=0}^{\infty} \frac{(-)^m (2m-1)!!}{(2x^2)^m}.
\end{align}
The polarizability at finite momentum $z>\sqrt{t}$ has the following series expansion
\begin{align}\label{eq:pi_zt_lowt_z_geq_sqrt_t}
    &{\tilde \Pi}(z,t\ll1) = 1 + \frac{\sqrt{\pi t}}{2 z} \mathrm{Li}_{\tfrac{1}{2}}\qty(\frac{e^{z^2/t}}{1 - e^{1/t}}) \nonumber \\
    &- \sum_{m=0}^{\infty} \frac{(-)^m (2m-1)!!}{2^{m+1}}  \mathrm{Li}_{m+1}\qty[\frac{1}{1 - e^{1/t}}] \qty(\frac{t}{z^2})^{m+1},
\end{align}
where $\mathrm{Li}_s(x) = \sum_{n=1}^{\infty} x^{n} n^{-s}$ is the polylogarithm function.
However, Eq.~\eqref{eq:pi_zt_lowt_z_geq_sqrt_t} has a convergence radius only at $z>\sqrt{t}$ and diverges at small $z \ll \sqrt{t}$.
A better analytical approximation which covers the full range of $z$ at $t < 1$ reads
\begin{align}\label{eq:pi_zt_lowt}
    &{\tilde \Pi}(z,t<1) = 1 + \frac{\sqrt{\pi t}}{2 z} \times \nonumber \\
    &\times \qty[\mathrm{Li}_{\tfrac{1}{2}}\qty(\frac{e^{z^2/t}}{1 - e^{1/t}}) - \mathrm{Li}_{\tfrac{1}{2}}\qty(\frac{1}{1 - e^{1/t}}) e^{-\frac{2z}{\sqrt{\pi t}}}].
\end{align}
For $t\ll1$, Eq.~\eqref{eq:pi_zt_lowt} also recovers the correct limiting values at $z\to 0$ [cf. Eq.~\eqref{eq:pi_z0}] and at $z \gg 1$ 
\begin{align}
    {\tilde \Pi}(z\gg 1,t\ll1) \approx 1 - \sqrt{1-z^{-2}},
\end{align}
where we use the large argument $x \to +\infty$ expansion of the polylogarithm function
\begin{align}\label{eq:polylog_large_argument}
    \mathrm{Li}_{s}(-e^x) = \sum_{k=0}^{\infty} \frac{B_{2k}}{(2k)!} \frac{(-)^k (1-2^{1-2k}) (2\pi)^{2k} x^{s-2k}}{\Gamma(s+1-2k)},
\end{align}
and $B_{2k}$ is the $2k$-th Bernoulli number.
With $\mathrm{Li}_{1/2}(-e^{x}) \approx -2\sqrt{x/\pi}$ as $x \to + \infty$, Eq.~\eqref{eq:pi_zt_lowt} recovers the correct zero-temperature polarizability as $t \to 0$. 
Moreover, Fig.~\ref{fig:pi_t} (a) shows that Eq.~\eqref{eq:pi_zt_lowt} is a good approximation of the low-temperature polarizability all the way through $t=1$.
For $z = 1$ and $t\ll1$, the argument of the polylogarithm function in Eq.~\eqref{eq:pi_zt_lowt} goes to $-1$ and $\mathrm{Li}_s(-1) = - (1 - 2^{1-s}) \zeta(s)$, where $\zeta(s)$ is the Riemann zeta function, so that to the leading order the polarizability becomes~\cite{Hwang_low_temp:2015}
\begin{align}\label{eq:pi_zt_lowt_z1}
    {\tilde \Pi}(z=1,t\ll1) \approx 1 - \frac{\sqrt{\pi t}}{2} (1 - \sqrt{2}) \zeta(1/2).
\end{align}
The full wave vector dependent polarizability, which includes the anomalous $\sqrt{T/E_F}$ suppression of screening around $q\approx 2k_F$, predicts a strong linear-in-$T$ increase of the metallic 2D resistivity at low temperatures~\cite{Stern:1980,Stern_screening:1985,Gold_Dolgopolov:1986,Hwang_low_T_R:1999,Hwang:2004}.
The significance of $2k_F$ screening in determining the 2D metallic temperature dependence in the disorder-limited resistivity was already pointed out by Stern a long time ago~\cite{Stern:1980}.
At high temperatures $t\gg 1$, the chemical potential $ \mu/T \to -\infty$ and we can expand the integrand of Eq.~\eqref{eq:pi_zt} using 
\begin{align}
    \frac{1}{4\cosh^2(x/2)} = -\sum \limits_{n=1}^{\infty} n(- e^{x})^n,\, x<0.
\end{align}
The series expansion of Eq.~\eqref{eq:pi_zt} at $t>1$ is given by
\begin{align}
    {\tilde \Pi}(z,t>1) = -\sum \limits_{n=1}^{\infty} (1 - e^{1/t})^n \sqrt{\frac{t}{n z^2}} F\qty(\sqrt{n z^2/t}), \label{eq:pi_zt_hight}
\end{align}
where $F(x)$ is the Dawson function defined as
\begin{align}
    F(x) = e^{-x^2} \int_0^{x} e^{u^2} du. 
\end{align}
Eq.~\eqref{eq:pi_zt_hight} has the following asymptotic limits. 
At $z\ll\sqrt{t}$ and $t > 1$, ${\tilde \Pi}(z,t) = 1 - e^{-1/t} +O(z^2/t^2)$; while at $z\gg\sqrt{t}$ and $t > 1$, ${\tilde \Pi}(z,t) = (2z^2)^{-1} +O(t/z^4)$.
If $t \gg 1$, Eq.~\eqref{eq:pi_zt_hight} is well approximated by 
\begin{align}
    {\tilde \Pi}(z,t\gg 1) = \frac{F\qty(z/\sqrt{t})}{z \sqrt{t}}. \label{eq:pi_zt_hight_fetter}
\end{align}
Eq.~\eqref{eq:pi_zt_hight_fetter} is the same result as obtained by Fetter~\cite{Fetter:1974,Ando_review:1982}, by identifying $q \lambda = z/\sqrt{t}$, where $\lambda = (2\pi \hbar^2/m T)^{1/2}$ is the thermal wavelength.
To the leading order at $z\ll\sqrt{t}$ and $t \gg 1$, the high-temperature polarizability ${\tilde \Pi} \approx 1/t$ is the well-known two-dimensional analog of the Debye-H\"{u}ckel screening~\cite{Fetter:1974,Ando_review:1982}.
Another useful analytical approximation to recover the correct $z=0$ value of the polarizability Eq.~\eqref{eq:pi_z0} at $t>1$ reads
\begin{align}
    {\tilde \Pi}(z,t> 1) = \frac{F\qty(z/\sqrt{t})}{z \sqrt{t}} + \qty(1 - e^{-1 / t} - \frac{1}{t}) e^{- \frac{\pi z^2}{2 t}}. \label{eq:pi_zt_hight_fullz}
\end{align}
Fig.~\ref{fig:pi_t} (b) shows Eq.~\eqref{eq:pi_zt_hight_fullz} is a good approximation of the high-temperature polarizability all the way through $t=1$.

Now we briefly discuss the Friedel oscillations associated with the polarizability behavior at $2k_F$.
At $T=0$, the kink in $\Pi(q)$ at $q=2k_F$ leads to an oscillatory term in the impurity potential in the real space, known as the Friedel oscillations~\cite{Friedel:1958,Stern:1967,Beal-Monod:1987,Litvinov:1998}.
The real space single-impurity potential is given by the Fourier transform
\begin{align}\label{eq:vr_fourier}
    U(r) = \int_0^{\infty} \frac{q dq}{2\pi} U(q) J_0(qr),
\end{align}
where we can expand $U(q)$ as
\begin{align}
    U(q) &= \frac{2\pi (e^2/\kappa) e^{-qd}}{q + q_{TF}\tilde \Pi(q/2k_F)} ,\\
    &= U_{TF}(q) \sum_{n=0}^{\infty} \qty(\frac{1-\tilde\Pi(q/2k_F)}{1 + q/q_{TF}})^n,\label{eq:vq_expansion}
\end{align}
The $n=0$ term in Eq.~\eqref{eq:vq_expansion} gives the conventional Thomas-Fermi screened potential, which is simply the long wavelength screening
\begin{align}
    U_{TF}(q) = \frac{2\pi e^2}{\kappa(q + q_{TF})}e^{-qd}.
\end{align}
In real space [cf. Eq.~\eqref{eq:vr_fourier}], the Thomas-Fermi screened potential can be well-approximated by the following analytical expression
\begin{align}
    U_{TF}(r) = \frac{e^2}{\kappa \sqrt{r^2 + d^2}} \frac{1+ q_{TF} d}{(1 + q_{TF} \sqrt{r^2 + d^2})^2}, \label{eq:V_TF(r)}
\end{align}
which gives the correct limits in the parameters $r/d$, $q_{TF} r$, and $q_{TF} d$. For example, in the limits of $r\gg d$ and $q_{TF} r \gg 1$, Eq.~\eqref{eq:V_TF(r)} reduces to the expression of the dipole potential given by Stern in Ref.~\cite{Stern:1967}, while Eq.~\eqref{eq:V_TF(r)} reduces to the point-charge Coulomb potential $U_{TF}(r) = e^2/\kappa \sqrt{r^2 + d^2}$ in the limits of weak screening $q_{TF} r, q_{TF}d \ll 1$.
Now we calculate the correction terms to $U_{TF}(r)$ which are the Friedel oscillations arising from the finite wavevector screening.
The leading correction is related to $n=1$ term in Eq.~\eqref{eq:vq_expansion}
\begin{align}
    &\Delta^{(1)} U(r) = \int_0^{\infty} \frac{q dq}{2\pi} J_0(qr) U_{TF}(q) \frac{1-\tilde\Pi(q/2k_F)}{1 + q/q_{TF}} , \\
    &\approx \frac{e^2 q_{TF} e^{-b}}{\kappa (1+s)^2} \int_1^{\infty} dz \sqrt{1-z^{-2}} J_0(2k_Fr z) e^{-b(z-1)},\label{eq:delta1_Vr}
\end{align}
where $z = q/2k_F$ [cf. Eq.~\eqref{eq:tilde_pi_t0}], $s = q_{TF}/2k_F$, and $b = 2k_F d$. 
We immediately see that the Friedel oscillation is exponentially suppressed in the case of a remote impurity far from the 2D system such that $b\gg 1$.
For a nearby impurity with $b \ll 1$, Eq.~\eqref{eq:delta1_Vr} is given by
\begin{align}
    \Delta^{(1)} U(r) &\approx \frac{e^2 q_{TF} e^{-b}}{\kappa (1+s)^2}\qty[\mathrm{Si}(\xi) + \frac{\cos\xi}{\xi} - \frac{\pi}{2}], \\
    &\approx - \frac{e^2 q_{TF} e^{-b}}{\kappa (1+s)^2} \qty[\frac{\sin\xi}{\xi^2} + O(\xi)^{-3}],\label{eq:delta1_Vr_large_r}
\end{align}
where $\mathrm{Si}(\xi) = \int_0^{\xi} (\sin x/x) dx$ is the sine integral function and $\xi = 2k_F r$. 
In the second step to obtain Eq.~\eqref{eq:delta1_Vr_large_r}, we employ the large $r$ limit such that $\xi = 2k_F r \gg 1$, and the large-argument expansion of $\mathrm{Si}(\xi)$
\begin{align}
    \mathrm{Si}(\xi\gg 1) = \frac{\pi}{2} - \frac{1}{\xi} \cos \xi - \frac{1}{\xi^2} \sin \xi + O(\xi)^{-3}.
\end{align}
Following the same procedure, we get the $n=2$ correction term in Eq.~\eqref{eq:vq_expansion} in the limit of $b \ll 1$ and $\xi \gg 1$
\begin{align}
    \Delta^{(2)} U(r) 
    &\approx - \frac{e^2 q_{TF} s e^{-b}}{\kappa (1+s)^3} \times\nonumber\\ 
    &\times \qty[\frac{2(\cos\xi + \sin \xi)}{\sqrt{\pi} \xi^{5/2}} + O(\xi)^{-7/2}].\label{eq:delta2_Vr_large_r}
\end{align}
Combining Eqs.~\eqref{eq:delta1_Vr_large_r} and \eqref{eq:delta2_Vr_large_r}, we obtain the same expression as given by Stern [cf. Eq. (11) in Ref.~\cite{Stern:1967}]:
\begin{align}
    \Delta U(r)& \approx - \frac{e^2 q_{TF} e^{-b}}{\kappa (1+s)^2} \times \nonumber \\
    &\times \qty[\frac{\sin\xi}{\xi^2} - \frac{2^{3/2} \cos(\xi - \pi/4)}{\sqrt{\pi}(1+s^{-1}) \xi^{5/2}} + O(\xi)^{-3}]. \label{eq:delta_Vr_Stern}
\end{align}
Higher-order corrections can be obtained systematically in a similar way.
Since the Friedel oscillations in Eq.~\eqref{eq:delta_Vr_Stern} correspond to Fourier components only for $q>2k_F$ while the largest $T=0$ backscattering momentum is $q=2k_F$, there is no scattering from Friedel oscillations at zero temperature. 
At low temperatures $T\ll E_F$, there is a small probability for an electron to have a momentum $k$ larger than $k_F$, so that the correction to the scattering cross section is proportional to $(k - k_F)/k_F \propto T/E_F$, leading to the linear-in-$T$ resistivity.
However, this is not the only contribution to the linear-in-$T$ resistivity.
The polarizability also acquires a temperature dependence which gives rise to the temperature-dependent Friedel oscillations~\cite{Zala_Aleiner:2001}
\begin{align}
    \Delta U(r,T) \approx - \frac{e^2 q_{TF} e^{-b}}{\kappa (1+s)^2} \frac{\sin (2k_F r)}{(2k_F r_t)^2\sinh^2\qty(r/r_t)},\label{eq:delta_Vrt}
\end{align}
where $r_t = \hbar v_F/T = 2/(tk_F)$.
At $r \ll r_t$, the denominator $r_t \sinh\qty(r/r_t) \to r$, and Eq.~\eqref{eq:delta_Vrt} reduces to Eq.~\eqref{eq:delta1_Vr_large_r}. 
While at $r \gg r_t$, $r_t \sinh\qty(r/r_t) \to r_t e^{r/r_T}/2$ and the Friedel oscillations are exponentially damped by a factor of $e^{-2r/r_t}$.
In the context of impurity scattering, $r_t$ has the physical meaning of the longest distance up to which the interference between the scattering from the impurity and from the Friedel oscillations can persist.
In this case, an electron with momentum $k$ can maintain phase coherence of the scattering from Friedel oscillations as long as $(k-k_F) \propto r_t^{-1} \sim (T/E_F)k_F$, again leading to the scattering cross section proportional to $T/E_F$.
A more detailed derivation of the coefficient of the linear-in-$T$ resistivity associated with the screened Coulomb impurity is given in Sec.~\ref{sec:t-dependent_resistivity}.

\section{Temperature dependent resistivity}
\label{sec:t-dependent_resistivity}
In this Section, we discuss the general numerical procedure to compute the temperature dependent resistivity and porvide some analytical results for the low- and high-temperature asymptotics.
We start with the Boltzmann kinetic equation, 
\begin{align}\label{eq:Boltzmann}
    \frac{\partial n}{\partial t} + \vb{v} \cdot \frac{\partial n}{\partial \vb{r}} + e\vb{E} \cdot \frac{\partial n}{\partial \vb{p}} = - \frac{n - n_0}{\tau},
\end{align}
where $n(\vb{r}, \vb{p})$ is the distribution function in the presence of an external electric field $\vb{E}$, and $n_0$ is the equilibrium distribution function without the electric field.
(We note that our notation $n$ is used for both the charge carrier density and the distribution function, although the distinction between them should be evident from the context.) 
$\tau$ is the scattering time. $\vb{v} = \vb{p}/m$ is the velocity.
In the steady state, which is spatially homogeneous, the first two terms of Eq.~\eqref{eq:Boltzmann} vanish.
Assuming a weak external electric field, we substitute $n = n_0 + n_1$ with a perturbation $n_1 \ll n_0$ and obtain the following
\begin{align}
    e\vb{E} \cdot \frac{\partial n_0}{\partial \vb{p}} = - \frac{n_1}{\tau}.
\end{align}
Since $n_0(\epsilon) = \qty[e^{(\epsilon - \mu)/T} + 1]^{-1}$ depends on the momentum only through the energy $\epsilon(\vb{p}) = p^2/2m$, we have $\partial n_0 / \partial \vb{p} = \vb{v} \cdot \partial n_0 / \partial \epsilon$ and the corresponding $n_1$ reads
\begin{align}
    n_1 = - e\vb{E} \cdot \vb{v} \tau \frac{\partial n_0}{\partial \epsilon}.
\end{align}
The electric current is expressed through the distribution function in the following manner
\begin{align}
    &\vb{j} = g e \int \vb{v} n \frac{d^d p}{(2\pi \hbar)^d} \\
    &= -e^2 \int \vb{v} (\vb{v} \cdot \vb{E}) \tau \frac{\partial n_0}{\partial \epsilon} \nu(\epsilon) d\epsilon \frac{d \Omega}{\Omega_0}
\end{align}
where $\nu(\epsilon)$ is the single-particle density of states (DOS), $g$ is the total quantum degeneracy (e.g. spin/valley degrees of freedom), $d$ is the spatial dimensionality, $d \Omega$ is the corresponding differential solid angle in $d$-dimensional space and $\Omega_0$ is the total solid angle. 
For example, if $d=2$, then $d \Omega = d \phi$ and $\Omega_0 = 2\pi$; while if $d = 3$, then $d \Omega = \sin \theta d\theta d \phi$ and $\Omega_0 = 4 \pi$.
The physical meaning of the integral $d \Omega/\Omega_0$ is to take the average of the angle between $\vb{p}$ and $\vb{E}$ and we have
\begin{align}
    \int \vb{v} (\vb{v} \cdot \vb{E}) \frac{d \Omega}{\Omega_0} = \frac{1}{d} v^2 \vb{E}.
\end{align}
The conductivity $\sigma$ defined through $\vb{j} = \sigma \vb{E}$ is given by
\begin{align}
    \sigma = e^2 \int D \nu(\epsilon) d\epsilon \qty(- \frac{\partial n_0}{\partial \epsilon}),
\end{align}
where we introduce the diffusion constant $D = v^2 \tau / d$.
At zero temperature, $- \partial n_0/\partial \epsilon = \delta(\epsilon - \mu)$ and the conductivity becomes $\sigma = e^2 [D \nu(\epsilon)]_{\epsilon = \mu}$, which is the well-known Einstein relation in the single-particle approximation.
As a side note, if we take into account the interaction between electrons, the Einstein's relation will incorporate the compressibility $\partial n/\partial \mu$ instead of the single-particle density of states. %It is important to note that the inclusion of the electron-electron interaction does not affect the compressibility.
Expressing the diffusion constant in terms of the energy $D = (\epsilon \tau / m) (2/d)$, we can rewrite the conductivity at finite temperatures through
\begin{align}\label{eq:conductivity}
    \sigma = \frac{n_e e^2 \tau_T}{m},
\end{align}
where $n_e$ is the electron density and $\tau_T$ is the energy-averaged scattering time at finite temperatures
\begin{align}\label{eq:tauT}
    \tau_T = \frac{\int_0^{\infty} \epsilon \tau(\epsilon) \nu(\epsilon) d\epsilon \qty(- \frac{\partial n_0}{\partial \epsilon})}{\int_0^{\infty} \epsilon \nu(\epsilon) d\epsilon \qty(- \frac{\partial n_0}{\partial \epsilon})}.
\end{align}
The temperature dependent mobility is $\mu_T = e \tau_T / m$.
The energy-average scattering rates of two different mechanisms $\tau_{T,1}^{-1}$ and $\tau_{T,2}^{-1}$ do not add in general, i.e., $\tau_T^{-1} \geq \tau_{T,1}^{-1} + \tau_{T,2}^{-1}$~\cite{Mansfield:1956,Ziman:2001}.
The denominator in Eq.~\eqref{eq:tauT} can be calculated explicitly
\begin{align}
    -\int \limits_0^{\infty} \epsilon \nu(\epsilon) d\epsilon \frac{\partial n_0}{\partial \epsilon} = \int \limits_0^{\infty} [\epsilon \nu'(\epsilon)  + \nu(\epsilon)]  n_0 d\epsilon = \frac{d}{2} n_e.
\end{align}
where we use $n_e = \int n_0 \nu(\epsilon) d\epsilon$ and the energy scaling of the $d$-dimensional DOS $\nu(\epsilon) \propto \epsilon^{(d/2) - 1}$ for a parabolic energy dispersion.
In the following, we focus on the case of interest where $d=2$.
The transport scattering time $\tau$ at energy $\epsilon = \hbar^2 k^2/2m$ can be evaluated using the Born approximation
\begin{align}\label{eq:1_tau}
	\frac{1}{\tau} =\frac{4m}{\pi \hbar^3} \int \limits_0^{2k}\frac{dq}{\sqrt{4k^2 - q^2}} \qty(\frac{q}{2k})^2 \ev{\abs{U(q)}^2}\,,
\end{align}
where $U(q)$ is the screened potential of a given scattering source.
For charged impurity scattering, the potential correlator averaged over impurity positions is given by 
\begin{align}\label{eq:uq2}
    \ev{\abs{U(q)}^2} = \int \limits_{-\infty}^{+\infty} dz \; N(z) U_1^2(q,T,z)\,,
\end{align}
where $N(z)$ is the 3D concentration of impurities at a distance $z$ from the center of the 2D system. 
$U_1(q,T,z)$ is the screened Coulomb potential for a single impurity located at $z$
\begin{align}\label{eq:u_i1_mu}
    U_1(q,T,z) = \frac{2\pi e^2}{\kappa q \epsilon(q,T) } e^{-q\abs{z}},
\end{align}
where the dielectric function is given by
\begin{align}\label{eq:df}
    \epsilon(q,T) = 1 + V(q) \Pi(q,T).
\end{align}
Here $V(q) = 2\pi e^2/\kappa q$ is the Coulomb interaction within the 2D system, $\Pi(q,T)$ is the static polarizability at finite temperatures defined through Eq.~\eqref{eq:polarization_def}.
The temperature dependence enters the calculation of $\tau_T$ through two parts.
The first temperature dependence is determined through the energy average weighted by the factor $(-\partial n_0/\partial \epsilon)$.
We note that this temperature dependence, arising entirely from the Fermi surface averaging, is always present, and contributes typically an increasing conductivity of $O(T/T_F)^2$ with increasing $T$ (which is the case for the 3D metal) simply because the effective scattering wavevector increases with increasing $T$, leading to a weaker Coulomb potential because of its $1/q$ dependence.
The second temperature dependence is due to the polarizability $\Pi(q,T)$ appearing in the screened potential, and this implies reduced screening at higher $T$ and $q\gtrsim 2k_F$, thus actually suppressing the conductivity, and giving an $O(T/T_F)$ contribution as discussed before.

Next, we present an example of the calculation of $\tau_T$ resulting from remote charged impurity scattering, where there is a layer of Coulomb impurities at distance $d$ from the 2D system and the impurity concentration is given by $N(z) = n_r \delta(z - d)$.
Introducing dimensionless quantities $s = q/2k_F$, $t = T/E_F$, $a = \epsilon/E_F$, $b = 2k_F d$, we have
\begin{align}\label{eq:1_tau_energy}
    \frac{1}{\tau(s,t,a,b)} = \frac{1}{\tau_0} f(s,t,a,b),
\end{align}
where 
\begin{align}
    \frac{1}{\tau_0} = n_r \frac{\pi^2 \hbar}{m} \qty(\frac{2}{g})^2,
\end{align}
and 
\begin{align}\label{eq:f_no_gate}
    f(s,t,a,b) = \frac{2}{\pi} \int \limits_0^1 \frac{2 x^2 e^{-2b\sqrt{a}x} dx}{\sqrt{1-x^2} \qty[x\sqrt{a}/s + \tilde{\Pi}(x\sqrt{a}, t)]^2},
\end{align}
where $\tilde{\Pi}(z,t)$ is the dimensionless polarizability defined through Eq.~\eqref{eq:pi_zt}.
Substituting Eq.~\eqref{eq:1_tau_energy} into Eq.~\eqref{eq:tauT} we obtain
\begin{align}\label{eq:tau_T_integral}
    \tau_T(s,t,b) = \tau_0 t \int \limits_0^{\infty} \frac{f(s,t,ut,b)^{-1} u du}{4 \cosh^2\qty[(u/2) - \ln(e^{1/t} - 1)/2]}
\end{align}
The zero-temperature scattering time is given by
\begin{align}
    \tau_T(s,t = 0,b) = \tau_0 f_0(s,b)^{-1},
\end{align}
where 
\begin{align}\label{eq:f0_no_gate}
    f_0(s,b) = \frac{2}{\pi} \int_{0}^{1} \frac{2x^2 e^{-2bx}\,dx}{\sqrt{1-x^2}(x/s+1)^2},
\end{align}
so that the zero-temperature resistivity reads
\begin{align}
    \rho_0 = \frac{m f_0(s,b)}{n_e e^2 \tau_0} = \frac{h}{e^2} \frac{n_r}{n_e} \qty(\frac{2}{g})^2 \frac{\pi}{2}f_0(s,b).
\end{align}
For in-plane impurities $b=0$, we have $f_0(s,b=0) = f_0(s)$, where 
\begin{align}
    f_0(s) &= 2s^2 + \frac{4 s^3}{\pi(1-s^2)} + \nonumber\\
    &+\begin{cases}
         \frac{4 s^3\left(2-s^2\right) \ln \qty[(1-\sqrt{1-s^2})/s] }{\pi\left(1-s^2\right)^{3/2}}, \, &s<1,\\
         \frac{4 s^3 (2-s^2) \sec^{-1} (s) }{\pi(s^2 - 1)^{3/2}}, \, &s\geq 1.
    \end{cases}\label{eq:f0s}
\end{align}
We have the limits $f_0(s\gg1) = 1$ and $f_0(s\ll1) = 2s^2$, which are well captured by the interpolation expression $f_0(s) \approx (1 + s^{-1}/\sqrt{2})^{-2}$.
For remote impurities far away the 2D system $b\gg 1$, we have $f_0(s\gg1, b\gg1) = (\pi b^3)^{-1}$.
The finite-temperature resistivity is given by
\begin{align}\label{eq:rho_t}
    \rho(t) = \rho_0 \frac{\tau_0}{f_0(s,b)\tau_T(s,t,b)}.
\end{align}
Next, we give some analytical results of the low- and high-temperature asymptotics of the resistivity for scattering from in-plane, remote, and background charged impurities, and charge-neutral atomic point defects.  
\subsection{in-plane charged impurity scattering}
\label{sec:in-plane_impurity_scattering}
\begin{figure}[t]
    \centering
    \includegraphics[width = \linewidth]{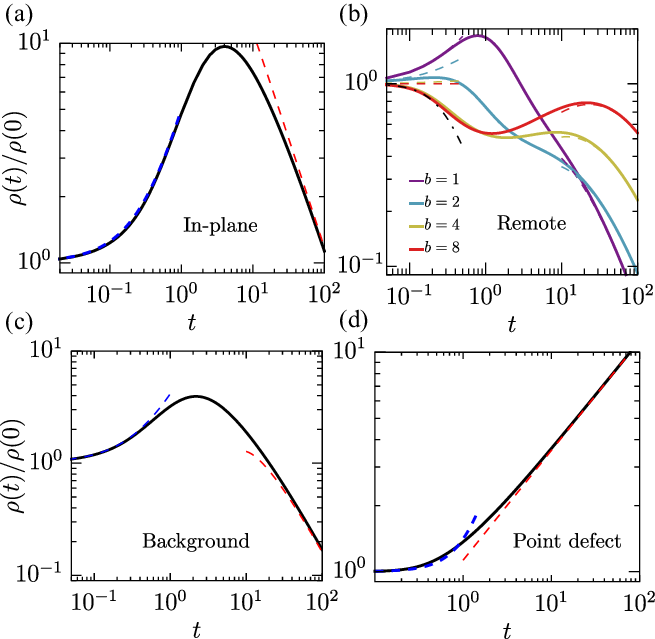}%mu_low_ne0.eps
    \caption{Resistivity $\rho(t)/\rho_0$ as a function of $t=T/E_F$ due to scattering from (a) in-plane charged impurities, (b) remote charged impurities, (c) background charged impurities, and (d) charge-neutral point defects. The solid curves are the numerical result calculated using (a) Eq.~\eqref{eq:rho_t} with $s=q_{TF}/2k_F=10$ and $b=2k_F d=0$, (b) Eq.~\eqref{eq:rho_t} with $s=10$ and $b=1,2,4,8$, (c) Eqs.~\eqref{eq:f_b_no_gate} and \eqref{eq:f_0b_no_gate} with $s=10$, and (d) Eq.~\eqref{eq:tau_Tv}. The dashed curves show the low- and high-temperature asymptotics (a) Eqs.~\eqref{eq:rho_tll1} and \eqref{eq:rho_tgg1}, (b) Eqs.~\eqref{eq:rho_low_t_analytic_small_b} and \eqref{eq:rho_high_t_analytic_large_b}, (c) Eqs.~\eqref{eq:rho_bg_low_t_analytic} and \eqref{eq:rho_bg_high_t_analytic}, and (d) Eqs.~\eqref{eq:rho_v_low_t} and \eqref{eq:rho_v_high_t} respectively. The black dot-dashed curve in (b) represents the low-temperature resistivity Eq.~\eqref{eq:rho_low_t_large_b} valid at $b \gg 1$, which is independent on the value of $b$. }
    \label{fig:rho_t_impurity}
\end{figure}

The low- and high-temperature resistivity for in-plane impurity scattering [cf. Eq.~\eqref{eq:rho_t} with $b = 2k_F d =0$] can be asymptotically expanded as~\cite{Hwang:2004,Hwang_low_temp:2015}
\begin{gather}\label{eq:rho_tll1}
    \rho(t \ll 1) = \rho_0 \qty[1 + \frac{2 s}{1+s}t + O(t)^{3/2}], \\
    \rho(t \gg 1) = \rho_0 s^2 t^{-1}\qty[1 + O(t)^{-3/2}].\label{eq:rho_tgg1}
\end{gather}
Figure~\ref{fig:rho_t_impurity} (a) compares the asymptotical expansions Eqs.~\eqref{eq:rho_tll1} and~\eqref{eq:rho_tgg1} with the numerical resistivity as a function of $t = T/E_F$ at $s=q_{TF}/2k_F = 10$ and $b=2 k_F d=0$.
In Ref.~\cite{Hwang:2004}, it was pointed out that the low-temperature expansion results of $\rho(t \ll 1)$ in Ref.~\cite{Gold_Dolgopolov:1986} have incorrect coefficients, and we confirm it here. 
Eq.~(2a) in Ref.~\cite{Hwang:2004} is the same as the Hartree term Eq.~(2.12) in Ref.~\cite{Zala_Aleiner:2001}.
Both are different from Eq.~(28) in Ref.~\cite{Gold_Dolgopolov:1986}, where the coefficient of the linear-in-$T$ term is larger by a factor of $2\ln 2$.
Here we present a way to fix the result of Ref.~\cite{Gold_Dolgopolov:1986} and obtain the correct linear-in-$T$ coefficient.
At low temperatures $t\ll 1$, the correction to the polarizability is small, $\Delta \tilde{\Pi}(z,t) = \tilde{\Pi}(z,t) - \tilde{\Pi}(z,0 ) \ll \tilde{\Pi}(z,0)$, such that one can expand $\tau^{-1}$ or the dimensionless scattering rate $f$ [cf. Eqs.~\eqref{eq:1_tau_energy} and \eqref{eq:f_no_gate}] in terms of $\Delta \tilde{\Pi}(z,t)$ as follows
\begin{align}
    &f(s,t,a) = \frac{2}{\pi} \int \limits_0^1 \frac{2 x^2 dx}{\sqrt{1-x^2} \qty[x\sqrt{a}/s + \tilde{\Pi}(x\sqrt{a}, 0)]^2} - \nonumber \\
    & - \frac{2}{\pi} \int \limits_0^1 \frac{2 x^2 dx}{\sqrt{1-x^2}} \frac{2 \Delta \tilde{\Pi}(x\sqrt{a},t)}{\qty[x\sqrt{a}/s + \tilde{\Pi}(x\sqrt{a}, 0)]^3} + O\qty(\Delta \tilde{\Pi})^{2} \nonumber \\
    &=f(s,t=0,a) + \Delta_t f(s,t,a).
\end{align}
We can further expand $f(s,t=0,a)$ around $a=1$ in terms of $\Delta \tilde{\Pi}(x\sqrt{a},0) = \tilde{\Pi}(x\sqrt{a},0) - \tilde{\Pi}(x,0) \ll \tilde{\Pi}(x,0)$, since in the end we should do the energy average of $\tau(\epsilon)$ using the Fermi-Dirac distribution, where the dominant contribution comes from $\abs{\epsilon/E_F-1} = \abs{a-1} \lesssim t \ll 1$:
\begin{align}
    &f(s,t=0,a) \nonumber \\
    &\approx \frac{2}{\pi} \int \limits_0^1 \frac{2 x^2 \qty[1 - \frac{2 \Delta \tilde{\Pi}(x\sqrt{a}, 0)}{x/s + \tilde{\Pi}(x, 0)}] dx}{\sqrt{1-x^2} \qty[x/s + \tilde{\Pi}(x, 0)]^2}  + O\qty(\Delta \tilde{\Pi})^{2}\nonumber \\
    &=f_0(s) + \Delta_a f(s,a).
\end{align}
As a result, at low temperatures $t\ll 1$, we can expand the dimensionless scattering rate as
\begin{align}
    f(s,t,a) &= f_0(s) + \Delta_a f(s,a) + \Delta_t f(s,t,a), \\
    &= f_0(s) + \sum_{n=1}^{\infty} \qty(\Delta_a^{(n)} f + \Delta_t^{(n)} f),
\end{align}
or the dimensionless scattering time as
\begin{align}
    f^{-1} &= f_0(s)^{-1} - f_0(s)^{-2} \sum_{n=1}^{\infty} \qty(\Delta_a^{(n)} f + \Delta_t^{(n)} f),
\end{align}
where $\Delta_a^{(n)} f$ and $\Delta_t^{(n)} f$ are given by
\begin{align}
    &\Delta_a^{(n)} f = \frac{2}{\pi} \int \limits_0^1 \frac{(2 x^2) (-)^n (n+1) \Delta \tilde{\Pi}(x\sqrt{a}, 0)^n dx}{\sqrt{1-x^2} \qty[x/s + \tilde{\Pi}(x, 0)]^{n+2}}, \label{eq:delta_a_f_n}\\
    &\Delta_t^{(n)} f = \frac{2}{\pi} \int \limits_0^1 \frac{(2 x^2) (-)^n (n+1) \Delta \tilde{\Pi}(x\sqrt{a},t)^n dx}{\sqrt{1-x^2} \qty[x/s + \tilde{\Pi}(x, 0)]^{n+2}} 
\end{align}
The corresponding corrections in the temperature dependent scattering time $\tau_T$ reads
\begin{align}
   \tau_T = \frac{\tau_0}{f_0(s)} - \frac{\tau_0}{f_0(s)^2}\sum_{n=1}^{\infty} \qty(\Delta_a^{(n)} f_T + \Delta_t^{(n)} f_T) \label{eq:tau_T_expansion}
\end{align}
where $\Delta_i^{(n)} f_T$ is the energy average of $\Delta_i^{(n)} f$ with $i = a,t$:
\begin{align}
    \Delta_i^{(n)} f_T &= \int_0^{\infty} \qty(\frac{\epsilon}{E_F}) \qty[\Delta_i^{(n)} f(\epsilon/E_F)] \qty(- \frac{\partial n_0}{\partial \epsilon}) d\epsilon, \\
    &=\int \limits_0^{\infty} \frac{ tu \qty[\Delta_i^{(n)} f(tu)] du}{4 \cosh^2\qty[(u/2) - \ln(e^{1/t} - 1)/2]}. \label{eq:delta_i_n_f_T}
\end{align}
First, we evaluate $\Delta_a^{(n)} f_T$ and show $n=1$ and $n=2$ correspond to $O(T)$ and $O(T^{3/2})$ contributions respectively.
Using the expression of $\Delta \tilde{\Pi}(x\sqrt{a}, 0)$ 
\begin{align}
    \Delta \tilde{\Pi}(x\sqrt{a}, 0) &= \sqrt{1-x^{-2}} \Theta(x-1) \nonumber \\
    &- \sqrt{1 - (x^2 a)^{-1}} \Theta(x - 1/\sqrt{a}),
\end{align}
$\Delta_a^{(n)} f$ can be rewritten as 
\begin{align}
    \Delta_a^{(n)} f &= \frac{8}{\pi} \int \limits_{1/\sqrt{a}}^1 \frac{x^2 (1 - 1/x^2 a)^{n/2} dx}{\sqrt{1-x^2} \qty(x/s + 1)^{(n+2)}} \Theta(a-1), \\
    &\approx \frac{8 N_n(a) \Theta(a-1)}{\pi \qty(1/s + 1)^{(n+2)}} .
\end{align}
In the second step, we use the fact that $0 < (a-1) \ll 1$ where $N_n(a)$ for a positive integer $n$ is given by
\begin{align}
    &N_n(a) = \int_{1/\sqrt{a}}^1 dx \frac{(x^2 - 1/a)^{n/2}}{\sqrt{1 - x^2}}, \\
    &= \frac{\sqrt{\pi} \Gamma\qty(\frac{n}{2} + 1)}{\Gamma \qty(\frac{n}{2} + \frac{3}{2})} (a-1)^{(n+1)/2} + O(a-1)^{(n+3)/2}.
    %- \frac{\sqrt{\pi} \Gamma\qty(\frac{n}{2} + 1) (n^2 + 4n +2)}{\Gamma \qty(\frac{n}{2} + \frac{5}{2})} (a-1)^{(n+3)/2}
\end{align}
For example, $N_1(a) = (\pi/4) (a-1) + O(a-1)^2$ and $N_2(a) = (2/3) (a-1)^{3/2} + O(a-1)^{5/2}$.
As a result,
\begin{align}\label{eq:delta_a_1_f}
    \Delta_a^{(1)} f &=  \Theta(a-1) \qty[\frac{2(a-1)}{(1/s + 1)^3} + O(a-1)^2], \\
    \Delta_a^{(2)} f &=  \Theta(a-1) \qty[ \frac{8}{\pi} \frac{(a-1)^{3/2}}{(1/s + 1)^4} + O(a-1)^{5/2}].\label{eq:delta_a_2_f}
\end{align}
Substituting Eqs.~\eqref{eq:delta_a_1_f} and \eqref{eq:delta_a_2_f} into Eq.~\eqref{eq:delta_i_n_f_T}, we obtain
\begin{align}
    \Delta_a^{(1)} f_T &= \frac{2\ln 2}{(1/s + 1)^3} t + O(t)^2, \label{eq:delta_a_1_f_T}\\
    \Delta_a^{(2)} f_T &= \frac{6\qty(1-1/\sqrt{2}) \zeta(3/2)}{\sqrt{\pi} (1/s + 1)^4} t^{3/2} + O(t)^{5/2}.\label{eq:delta_a_2_f_T}
\end{align}
Next, we evaluate $\Delta_t^{(n)} f_T$ and show that $n=1$ and $n=2$ correspond to another $O(T)$ and $O(T^{3/2})$ contributions respectively.
Using the expression of $\Delta \tilde{\Pi}(x\sqrt{a}, t)$ 
\begin{align}
    &\Delta \tilde{\Pi}(x\sqrt{a}, t) = \frac{\sqrt{\pi t}}{2 x\sqrt{a}} \mathrm{Li}_{\tfrac{1}{2}}\qty(\frac{e^{x^2 a/t}}{1 - e^{1/t}}) \nonumber \\
    &+ \Theta(x - 1/\sqrt{a}) \sqrt{1- \frac{1}{x^2 a}} .
\end{align}
For $a \leq 1$ and $t \ll 1$, we have 
\begin{align}
    \Delta_t^{(n)} f &=  \frac{4 (n+1)}{\pi} \int \limits_0^1  \frac{x^2 \qty[-\frac{\sqrt{\pi t}}{2 x\sqrt{a}}\mathrm{Li}_{1/2}\qty(-e^{(x^2 a - 1)/t})]^n dx}{\sqrt{1-x^2} \qty(x/s + 1)^{n+2}}, \nonumber \\
    &\approx \Delta_t^{(n)} f(a=1) \qty[\frac{\mathrm{Li}_{1/2}\qty(-e^{-\abs{1-a}/t})}{\mathrm{Li}_{1/2}\qty(-1)}]^n, \label{eq:delta_t_f_n}
\end{align}
Therefore, 
\begin{align}
    \Delta_t^{(n)} f_T \approx L_n \Delta_t^{(n)} f(a=1), \label{eq:delta_t_f_T_n}
\end{align}
where 
\begin{align}
    L_n = \int \limits_{-\infty}^{+\infty} dx \frac{\qty[\mathrm{Li}_{1/2}\qty(-e^{-\abs{x}})]^n}{4 \cosh(x/2)^2 \qty[\mathrm{Li}_{1/2}\qty(-1)]^n}.
\end{align}
For example, $L_1 = 0.4655$ and $L_2 = 0.2989$.
From Eq.~\eqref{eq:delta_t_f_n} we see that $\Delta_t^{(n)} f(a)$ peaks at $a=1$ and exponentially decay to zero with a small ``decay energy'' equal to $t$.
The peak value is given by
\begin{align}
    &\Delta_t^{(n)} f(a=1) \nonumber \\
    %&= \frac{4 (n+1)}{\pi} \int \limits_0^1  \frac{x^2 \qty[- \frac{\sqrt{\pi t}}{2 x}\mathrm{Li}_{1/2}\qty(-e^{(x^2-1)/t})]^n dx}{\sqrt{1-x^2} \qty(x/s + 1)^{n+2}}, \nonumber \\
    &\approx \frac{(n+1) \pi^{n/2} t^{(n+1)/2}}{\pi 2^{n-1} (1/s + 1)^{n+2}} \int \limits_0^{\infty}  \frac{du}{\sqrt{u}}\qty[-\mathrm{Li}_{1/2}\qty(-e^{-u})]^n, \nonumber \\
    &= \frac{(n+1) (\pi t)^{(n+1)/2}}{\pi 2^{n-1} (1/s + 1)^{n+2}} M_n,
    %&\approx \frac{4 (n+1) \pi^{n/2} t^{(n+1)/2}}{ \pi 2^{n+1} (1/s + 1)^{n+2}} \int \limits_0^{\infty}  \frac{\qty[-\mathrm{Li}_{1/2}\qty(-e^{(x^2-1)/t})]^n dx}{\sqrt{1-x^2}}
\end{align}
where 
\begin{align}
    M_n = (-)^n \sum_{\{m_k\}} \frac{(-)^{\sum_k m_k}}{\sqrt{\qty(\prod_k m_k) \qty(\sum_k m_k)}},
\end{align}
and the dummy indices $m_k$ sum over all positive integers, and the subscript $k$ ranges from 1 to $n$. For example, 
\begin{gather}
    M_1 = - \sum_{m=1}^{\infty} \frac{(-)^m}{m} = \ln 2, \\
    M_2 = \sum_{m_1=1}^{\infty}\sum_{m_2=1}^{\infty} \frac{(-)^{m_1+m_2}}{\sqrt{m_1^2 m_2 + m_1 m_2^2}} \approx 0.3066 ,
\end{gather}
we have
\begin{align}
    &\Delta_t^{(1)} f(a=1) = t \frac{2\ln 2}{(1/s + 1)^{3}}, \\
    &\Delta_t^{(2)} f(a=1) = t^{3/2} \frac{0.8150}{(1/s + 1)^{4}}.
\end{align}
Substituting them into Eq.~\eqref{eq:delta_t_f_T_n}, we obtain
\begin{align}
    &\Delta_t^{(1)} f_T = t \frac{0.6453}{(1/s + 1)^{3}}, \label{eq:delta_t_1_f_T}\\
    &\Delta_t^{(2)} f_T = t^{3/2} \frac{0.2436}{(1/s + 1)^{4}}. \label{eq:delta_t_2_f_T}
\end{align}
Combining Eqs.~\eqref{eq:rho_t}, \eqref{eq:tau_T_expansion}, \eqref{eq:delta_a_1_f_T}, \eqref{eq:delta_a_2_f_T}, \eqref{eq:delta_t_1_f_T}, and \eqref{eq:delta_t_2_f_T}, we finally obtain 
\begin{align}\label{eq:rho_low_t_analytic}
    \frac{\rho(t\ll1)}{\rho_0} \approx 1 + \frac{2.031 \, t}{(1/s + 1)^{3} f_0(s)} +  \frac{2.834\, t^{3/2}}{(1/s + 1)^{4} f_0(s)},
\end{align}
whose numerical coefficients are close to Eq.~\eqref{eq:rho_tll1} within a few percents difference, if we identify $f_0(s) \approx (1 + 1/s)^{-2}$ for $s \gg 1$.
On the other hand, in the high-temperature limit $t \gg 1$, the polarizability $\tilde \Pi \approx 1/t$ and we obtain
\begin{align}
    f(s,t,a) \approx t^2 f_0(s/t\sqrt{a}),
\end{align}
where $f_0(x)$ is given by Eq.~\eqref{eq:f0s}.
For example, the leading term in $t\gg1$ is $f_0(s/t\sqrt{a} \to 0) \approx 2s^2/t^2a$ and $f(s,t\gg 1,a) \approx 2 s^2/a$.
Together with the large-$t$ limit of Eq.~\eqref{eq:tau_T_integral}:
\begin{align}
    \tau_T \approx \tau_0 \int_0^{\infty} e^{-u} u f^{-1}(s,t, ut) du,
\end{align}
we obtain $\tau_T = \tau_0 t/s^2$ and $\rho(t\gg1) \approx \rho_0 s^2/t$.
Following the same procedure for the higher order terms we obtain Eq.~\eqref{eq:rho_tgg1}
\begin{gather}
    \rho(t \gg 1) \approx \rho_0 \frac{s^2}{t}\qty[1 -\frac{3 s t^{-3/2}}{4\sqrt{\pi}} \qty(1.09 + \ln\frac{t^3}{s^2}) ].
\end{gather}
%\YH{Add derivation of the high-temperature limit.}
\subsection{remote charged impurity scattering}
For remote impurities whose location $z=d$ is not too far from the 2D system such that $b=2k_F d \lesssim 1$, Eq.~\eqref{eq:rho_low_t_analytic} becomes
\begin{align}
    \frac{\rho(t\ll 1)}{\rho_0} &\approx 1 + \frac{2.031 e^{-2b}\, t}{(s^{-1} + 1)^{3} f_0(s,b)} + \frac{2.834 e^{-2b}\, t^{3/2}}{(s^{-1} + 1)^{4} f_0(s,b)}\nonumber \\
    &  + \qty(1-b-\frac{7\pi}{8})\frac{\pi^2}{12}\frac{e^{-2b}\, t^2}{(s^{-1} + 1)^{3} f_0(s,b)}. \label{eq:rho_low_t_analytic_small_b}
\end{align}
Since the linear-in-$T$ temperature dependence arises from $2k_F$-screening, but the remote impurities are ineffectively screened, so the effect is suppressed by an exponential factor $e^{-2b}$ as shown in Eq.~\eqref{eq:rho_low_t_analytic_small_b}.  
%Such suppression of linear-in-$T$ temperature dependence is dominant when the nearby impurities dominate the measured conductivity, which is often the case even when the system is modulation doped.
However, if remote impurities are very far from the 2D system such that $b = 2k_F d \gg 1$, the above result will be dramatically modified so that the resistivity decreases as the temperature increases, in opposite to the case $b \lesssim 1$ where impurities are close to the 2D system.
The reason, essentially a thermal averaging effect, for such a qualitative change of temperature dependence is as follows.
From Eq.~\eqref{eq:f_no_gate} we see that due to the exponential suppression of the remote impurity potential, the relevant momentum is at $x = q/2k \sim (2b\sqrt{a})^{-1}$, or $q/2k_F = x \sqrt{a} \sim (2b)^{-1} \ll 1$. 
Therefore, the strong low-temperature suppression of screening at $q \gtrsim 2k_F$ is irrelevant if $b\gg 1$, and we obtain
\begin{align}
    f(s,t\ll1, a, b\gg1) \approx \frac{1}{\pi b^3 a^{3/2}},
\end{align}
Substituting it into Eq.~\eqref{eq:tauT} and performing the integral using Eq.~\eqref{eq:cosh_expansion_low_T}, we get
\begin{align}
    \tau_T \approx \tau_0 \pi b^3 \Gamma(7/2) t^{5/2} (-) \mathrm{Li}_{5/2}(1-e^{1/t}).
\end{align}
Using the asymptotic expansion of the polylogarithm function Eq.~\eqref{eq:polylog_large_argument}, we get the resistivity at low temperatures in the limit $b \gg 1$:
\begin{align}\label{eq:rho_low_t_large_b}
    \frac{\rho(t\ll1)}{\rho_0} &\approx  \qty[\Gamma(7/2) t^{5/2} (-) \mathrm{Li}_{5/2}(1-e^{1/t})]^{-1}, \\
    &\approx 1 - \frac{35 \pi^2}{48} t^2 + O(t)^4,
\end{align}
which is independent on $b$.
Eq.~\eqref{eq:rho_low_t_large_b} is the same as Eq.~(B19) in Ref.~\cite{Lavasani:2019} up to a numerical factor.
On the other hand, at high temperatures $t \gg 1$, the polarizability $\tilde{\Pi} \approx 1/t$, and the relevant energy scale in the integral of $\tau_T$ is $\epsilon \sim T$ or $a = \epsilon/E_F \sim t \gg 1$, so that the relevant momentum $x \sim (2b\sqrt{a})^{-1} \ll 1$, and we obtain
\begin{align}\label{eq:rate_high_t_large_b}
    f(s,t,a,b) \approx \frac{2}{\pi} \int \limits_0^1 \frac{2 x^2 e^{-2b\sqrt{a}x} dx}{\qty(x\sqrt{a}/s + 1/t)^2} \approx \frac{f_{a1}(s,t,b)}{a^{3/2}},
\end{align}
where
\begin{align}
    &f_{a1}(s,t,b) = f(s,t,a=1,b) \nonumber \\
    &\approx \frac{2 s^2}{\pi}\qty[\frac{2 b s + t}{b t} + \frac{4s(b s + t) e^{2bs/t} \mathrm{Ei}\qty(-\tfrac{2bs}{t})}{t^2}],
\end{align}
and $\mathrm{Ei}(x) = -\int_{-x}^{\infty} (e^{-t}/t) dt$ is the exponential integral function.
Substituting Eq.~\eqref{eq:rate_high_t_large_b} into Eqs.~\eqref{eq:tauT} and \eqref{eq:rho_t} we obtain 
\begin{align}
    \tau_T \approx \frac{\tau_0 \Gamma(7/2) t^{3/2}}{f_{a1}(s,t,b)},
\end{align}
and
\begin{align}\label{eq:rho_high_t_analytic_large_b}
    \frac{\rho(t\gg 1)}{\rho(0)} \approx  \frac{f_{a1}(s,t,b)}{f_0(s,b) \Gamma(7/2) t^{3/2}},
\end{align}
The comparison of the numerical results of $\rho(t)$ with low- and high-$T$ asymptotics at different values of $b=1,2,4,8$ is shown in Fig.~\ref{fig:rho_t_impurity} (b).

\subsection{background charged impurity scattering}
For uniform background impurity concentration $N(z) = N_b$, the potential correlator is given by [cf. Eq.~\eqref{eq:uq2}]
\begin{align}
    \ev{\abs{U(q,T)}^2} = \frac{N_b}{q} \qty[\frac{2\pi e^2}{\kappa q \epsilon(q,T)}]^2,
\end{align}
Introducing the effective 2D impurity concentration $n_b = N_b / 2k_F$, the scattering rate [cf. Eqs.~\eqref{eq:f_no_gate} and~\eqref{eq:f0_no_gate}] for background impurity scattering $\tau_b^{-1} = \tau_0^{-1}(n_b) f_b$ can be rewritten as
\begin{align}\label{eq:f_b_no_gate}
    f_{b}(s,t,a) = \frac{2}{\pi} \int \limits_0^1 \frac{2 (x/\sqrt{a}) dx}{\sqrt{1-x^2} \qty[x \sqrt{a}/s + {\tilde \Pi}(x\sqrt{a}, t)]^2}.
\end{align}
At zero temperature, $\tau_b^{-1} = \tau_0^{-1}(n_b) f_{0b}$ where
\begin{align}\label{eq:f_0b_no_gate}
    f_{0b}(s) &= \frac{2}{\pi} \int \limits_0^1 \frac{2 x dx}{\sqrt{1-x^2} (x/s + 1)^2}, \\
    &= \frac{4}{\pi} 
    \begin{cases}
        \frac{s^2}{s^2 - 1} - \frac{s^2 \mathrm{sec}^{-1}(s)}{(s^2 - 1)^{3/2}}, \, &s\geq 1, \\
        \frac{s^2}{s^2 - 1} + \frac{s^2 \ln(\sqrt{s^{-2} - 1} + s^{-1})}{(1 - s^2)^{3/2}}, \, &s<1.
    \end{cases}
\end{align}
If $s \gg 1$, we have $f_{0b}(s) \approx (4/\pi) (1/s +1)^{-2}$.
At low temperatures $t\ll 1$, the most important momentum scale for background impurity scattering is $q \simeq 2k_F$ similar to the case for the in-plane impurity scattering discussed in Sec.~\ref{sec:in-plane_impurity_scattering}.
Therefore, following similar procedure we obtain the low-temperature asymptotic expansion of the resistivity
\begin{align}\label{eq:rho_bg_low_t_analytic}
    &\rho(t\ll 1)/\rho_0  \nonumber \\
    &\approx 1 + \frac{2.031 \, t}{(1/s + 1)^{3} f_{0b}(s)} +  \frac{2.834\, t^{3/2}}{(1/s + 1)^{4} f_{0b}(s)}.
\end{align}
At high temperatures, ${\tilde \Pi}(x\sqrt{a}, t) \sim 1/t$, and $a \sim t \gg 1$, such that the scattering rate becomes
\begin{align}
    f_{b}(s,t,a) &\approx \frac{2}{\pi} \int \limits_0^1 \frac{2 x dx}{\sqrt{1-x^2} \qty(x \sqrt{a}/s + 1/t)^2} \\
    &= \frac{s^2}{a^{3/2}} f_{0b}(s/t \sqrt{a}) \approx \frac{4}{\pi} \frac{s^2}{a^{3/2}} \ln\qty(\frac{t\sqrt{a}}{s}),
\end{align}
where we use $f_{0b}(x\ll1) \approx (4/\pi) \ln(1/x)$.
As a result, we obtain the high-temperature asymptotic expansion of the resistivity
\begin{align}\label{eq:rho_bg_high_t_analytic}
    \frac{\rho(t\gg 1)}{\rho_0} \approx \frac{4 s^2 \ln\qty(t^{3/2}/s)}{\pi \Gamma(7/2) f_{0b}(s)}  t^{-3/2}.
\end{align}
The comparison of the numerical results of $\rho(t)$ with low- and high-$T$ asymptotics is shown in Fig.~\ref{fig:rho_t_impurity} (c).

\subsection{charge-neutral atomic point defect scattering}
The scanning tunneling microscopy (STM) images reported in Ref.~\cite{liu2023twostep} explicitly show short-range intrinsic charge-neutral point defects in monolayer WSe$_2$. (Most likely Se vacancies.)
The density of charged and isovalent (charge neutral) point defects is of the order $3\times 10^9$ and $8\times 10^{10}$ cm$^{-2}$, respectively, for a flux-grown Se:W ratio of 100:1.
In addition, theoretical density functional theory (DFT) calculations of mid-gap states induced by point defects at either the W or Se sites also show a limited density of states (DOS) for Se vacancies~\cite{Kaasbjerg:2020}.
Since the charge carriers are entirely repelled by the depletion region around the atomic point defects such as Se vacancies, the scattering from the depletion region cannot be described by the Born approximation, but rather should be described by a hard-core scattering.
The scattering cross section $\Sigma$ is given by
\begin{align}
    \Sigma = \frac{1}{n_v l},
\end{align}
where $n_v$ is the density of the vacancies and $l = v_F \tau$ is the mean free path.
The corresponding scattering rate reads
\begin{align}\label{eq:1_tau_v}
    \frac{1}{\tau} = n_v v_F \Sigma = \frac{1}{\tau_0(n_v)} \frac{k_F \Sigma}{\pi^2} \qty(\frac{g}{2})^2,
\end{align}
where $\tau_0^{-1}(n_v) = n_v (2/g)^2 \pi^2 \hbar / m$.
The Drude conductivity can be rewritten in terms of the scattering cross section
\begin{align}
    \sigma = \frac{n e^2 \tau}{m } = \frac{n e^2}{ m  v_F n_v \Sigma},
\end{align}
and the corresponding mobility is given by
\begin{align}\label{eq:mobility_v}
    \mu_{\mathrm{PD}} = \sigma/e n = \frac{e}{\hbar k_F n \Sigma}.
\end{align}
We see that the mobility decreases as the density increases.
In both classical and quantum mechanics, the scattering cross section can be analytically evaluated and $\Sigma = (8/3) a_0$~\cite{Foulk:2021}, where $a_0$ is the radius of the atomic point defects and is roughly equal to the lattice constant $\sim 0.33$ nm of WSe$_2$. 
Using Eqs.~\eqref{eq:tauT} and~\eqref{eq:1_tau_v} and introducing a dimensionless quantity $c = k_F \Sigma$, we get the temperature-dependent scattering time contributed from the atomic vacancies
\begin{align}\label{eq:tau_Tv}
    \tau_{T}(t, c) =  \int \limits_0^{\infty} \frac{\tau_0(n_v) t f_v(c\sqrt{u t})^{-1} u du}{4 \cosh^2\qty[(u/2) - \ln(e^{1/t} - 1)/2]},
\end{align}
where 
\begin{align}
    f_v(c) = \frac{c}{\pi^2} \qty(\frac{g}{2})^2.
\end{align}
Equation~\eqref{eq:tau_Tv} can be evaluated analytically and we obtain
\begin{align}\label{eq:tau_Tv_analytical}
    \frac{\tau_{T}(t, c)}{\tau_0(n_v)} = \pi^2 \qty(\frac{2}{g})^2 \frac{\sqrt{\pi t}}{2 c}(-) \mathrm{Li}_{1/2}(1 - e^{1/t}).
\end{align}
As $t \to 0$, we have $(-)\mathrm{Li}_{1/2}(1 - e^{1/t}) \to (2/\sqrt{\pi t})$, and Eq.~\eqref{eq:tau_Tv_analytical} recovers the zero-temperature result Eq.~\eqref{eq:1_tau_v}.
At high temperatures $t \to \infty$, we have $(-)\mathrm{Li}_{1/2}(1 - e^{1/t}) \to 1/t$ and 
\begin{align}\label{eq:rho_v_high_t}
    \rho(t\gg 1)/\rho_0 \approx  2\sqrt{t/\pi}.
\end{align}
Using Eq.~\eqref{eq:polylog_large_argument}, we get the resistivity at low temperatures
\begin{align}\label{eq:rho_v_low_t}
    \frac{\rho(t\ll 1)}{\rho_0} &\approx  \qty[\Gamma(3/2) t^{1/2} (-) \mathrm{Li}_{1/2}(1-e^{1/t})]^{-1} \\
    & \approx 1 + \frac{\pi^2}{24}t^2.
\end{align}
The low- and high-$T$ asymptotics are shown in Fig.~\ref{fig:rho_t_impurity} (d).

\section{Double-gate screening}
\label{sec:gate}
We consider the effect of double-gate screening in this section.
The distance from the top (bottom) gate to the 2D system is $d_1$ ($d_2$).
The single-impurity Coulomb potential in the presence of double-gate screening is given by
\begin{align}
    U_1(q,T,z) = \frac{2\pi e^2}{\kappa q \epsilon(q,T)} D(qz,qd_1,qd_2),
\end{align}
where 
\begin{align}
    &D(qz,qd_1,qd_2) = e^{-q\abs{z}} -\nonumber \\
    &- \frac{e^{q z} (e^{2q(d_1 -z)} - 1) + e^{-q z} (e^{2q(d_2 +z)} - 1)}{e^{2q(d_1+d_2)} - 1},\label{eq:Dz}
\end{align}
The dielectric function is given by Eq.~\eqref{eq:df} where the bare Coulomb interaction within the 2D system is also screened by the double gates $V(q) = (2\pi e^2/\kappa q)D(0,qd_1, qd_2)$, where
\begin{align}
   &D(0,qd_1, qd_2) =  \tanh[q (d_1 + d_2)/2] + \nonumber \\
   &+ \frac{1 - \coth[q (d_1 + d_2)]}{2} \qty(e^{q d_1} - e^{q d_2})^2. \label{eq:D0}
\end{align}
For symmetric gates such that $d_1 = d_2 = d_g$, we have $V(q) = (2\pi e^2/\kappa q) \tanh(q d_g)$.
If $qd_1, qd_2 \gg 1$, then $D(0,qd_1, qd_2) \to 1$; while if $qd_1, qd_2 \ll 1$, then $D(0,qd_1, qd_2) \to 2 q d_1 d_2/(d_1 + d_2)$.
The double gate screening changes the remote-impurity scattering by modifying Eqs.~\eqref{eq:f_no_gate} and~\eqref{eq:f0_no_gate} as
\begin{align}
    &f(s,t,a,b,b_1,b_2) =  \nonumber \\
    &\frac{2}{\pi} \int \limits_0^1 \frac{2 x^2 (1-x^2)^{-1/2} D(x \sqrt{a} b,x\sqrt{a} b_1,x\sqrt{a} b_2)^2 dx}{\qty[\frac{x\sqrt{a}}{s} + \tilde{\Pi}(x\sqrt{a}, t) D(0,x\sqrt{a} b_1,x\sqrt{a} b_2)]^2},\label{eq:f_remote_gate}
\end{align}
and 
\begin{align}
    &f_0(s,b,b_1,b_2) = \frac{2}{\pi} \int \limits_0^1 \frac{2 x^2 D(x b,x b_1,x b_2)^2 dx}{\sqrt{1-x^2} \qty[\frac{x}{s} + D(0,x b_1,x b_2)]^2},\label{eq:f0_remote_gate}
\end{align}
where $b_1 = 2k_F d_1$ and $b_2 = 2k_F d_2$. 
If the gates are far away such that $b_1,b_2 \to +\infty$, then $D(x b,x b_1,x b_2) \approx e^{-2bx}$ and $D(0,x b_1,x b_2) \approx 1$. 
As a result, the remote impurity scattering results Eqs.~\eqref{eq:f_remote_gate} and \eqref{eq:f0_remote_gate} reduce back to the results without gate screening Eqs.~\eqref{eq:f_no_gate} and~\eqref{eq:f0_no_gate}.
If the delta-layer remote impurities are very close to the gate with a separation distance $d_0 \ll d_1,d_2$, we can further simplify Eqs.~\eqref{eq:Dz}.
For example, for RIs located at $z=d_1 - d_0$, we have
\begin{align}
    D(qd_1 - qd_0, q d_1, q d_2) \approx  
    \begin{cases}
        2 q d_0 e^{-q d_1}, \, &qd_1, qd_2 \gg 1, \\
        2 q d_0 \frac{d_2}{d_1 + d_2}, \, &qd_1, qd_2 \ll 1.
    \end{cases}
\end{align}
As a result, the transport scattering rate for RIs near the top gate reads
\begin{align}\label{eq:rate_RI_near_gate}
    f_0(s,b_1 - b_0,b_1,b_2) \approx 
    \begin{cases}
        \frac{3(2d_0)^2}{d_1^2} \frac{1}{\pi b_1^3}, \, &b_1, b_2 \gg 1, \\
        \frac{(2d_0)^2 d_2^2}{[q_{TF}^{-1}(d_1 + d_2) + 2d_1 d_2]^2}, \, &b_1, b_2 \ll 1.
    \end{cases}
\end{align}
In the limits of $b_1, b_2 \gg 1$ (i.e. $k_F d_1, k_F d_2 \gg 1$), if we compare the result Eq.~\eqref{eq:rate_RI_near_gate} with the conventional RI scattering rate expression in the absence of gate screening where 
\begin{align}\label{eq:rate_RI_no_gate}
    \frac{1}{\tau_{\mathrm{RI}}} = n_r \qty(\frac{2}{g})^2 \frac{\pi\hbar}{8m (k_F d_1)^3} = \frac{1}{\tau_0(n_r)} \frac{1}{\pi b_1^3}, \, b_1 \gg 1, 
\end{align}
we find the RI scattering rate is reduced by a factor of $3 (2d_0)^2/d_1^2$ due to the screening from the image charges inside the gate.
On the other hand, in the opposite limits when the carrier density is low $k_F d_1, k_F d_2 \ll 1$, RIs behave as in-plane charged impurities (i.e. $\tau^{-1}_{\mathrm{RI}}$ becomes independent on $k_F$ at small $k_F$) with an extra geometrical factor as defined in Eq.~\eqref{eq:rate_RI_near_gate}.
\begin{figure*}[t]
    \centering
    \includegraphics[width = 0.8\linewidth]{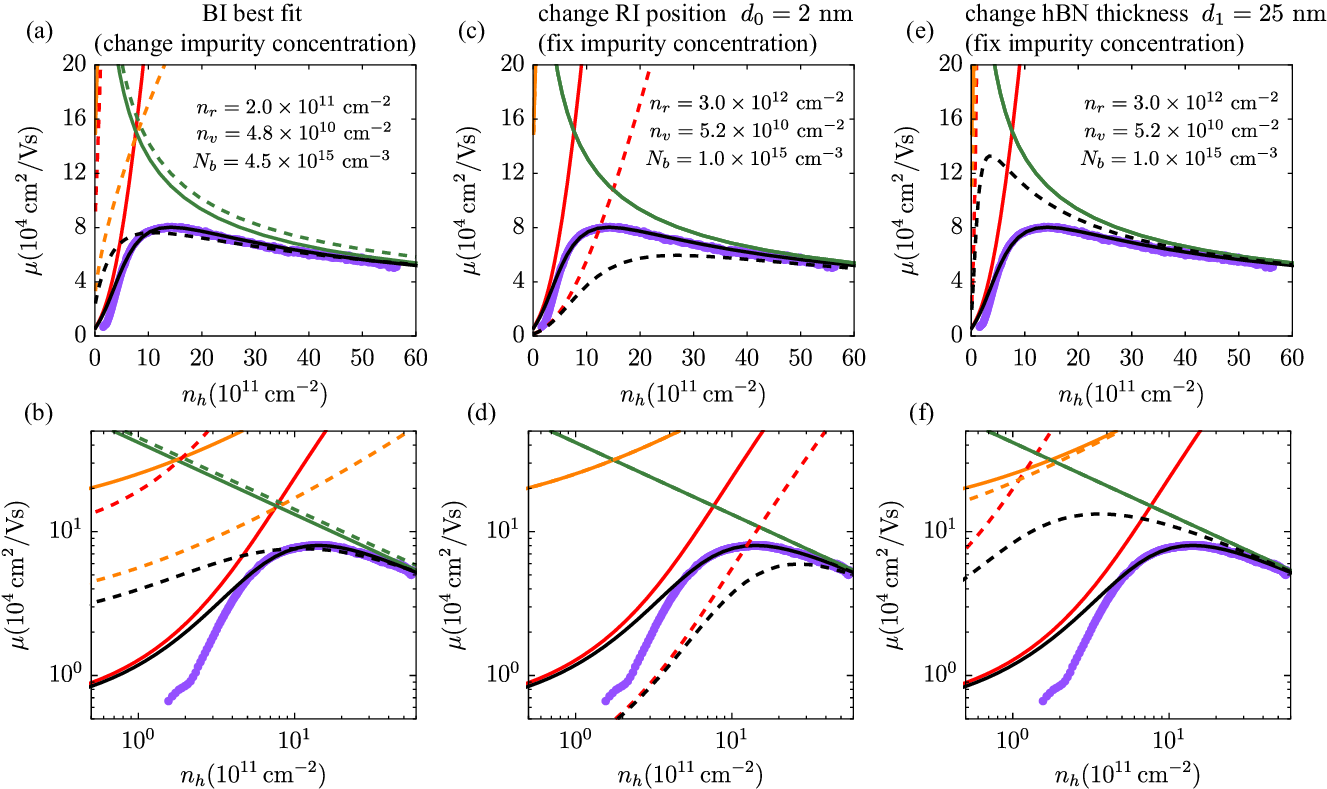}
    \caption{Mobility $\mu$ versus hole density $n_h$ in a monolayer WSe$_2$. (a) (c) (e) show the results in linear-linear scale, while (b) (d) (f) show the corresponding results in log-log scale. Purple dots are experimental data measured at $T = 1.5$ K. Solid curves represent the best-fit results, the same as shown in Fig.~\ref{fig:muT_n} (c), where RI is the dominant scattering mechanism at low densities. Dashed curves show different results described by the titles above the figures, respectively. The black curve represents the total mobility. Red (orange) curves represent the contribution from RI (BI) scattering. Green curves represent the PD scattering contribution.}
    \label{fig:mobility_nh}
\end{figure*}

For BI scattering, the potential correlator is given by Eq.~\eqref{eq:uq2} with 3D impurity concentration $N(z) = N_b [\Theta(d_1 - z) - \Theta(z + d_2)]$
\begin{align}
    \ev{\abs{U}^2} = \frac{N_b}{q} \qty[\frac{2\pi e^2}{\kappa q \epsilon(q,T)}]^2 D_b(qd_1,qd_2),
\end{align}
where the form factor $D_b(qd_1,qd_2)$ is given by
\begin{align}
    &D_b(qd_1,qd_2) = \int_{-d_2}^{d_1} q dz  D(qz,qd_1,qd_2)^2,  \\
    = &\mathrm{csch}(qd_1 + qd_2)^2 \left[\qty(\sinh(2 qd_1) - 2qd_1) \sinh(qd_2)^2 \right. \nonumber \\
    &\left.+ (\sinh(2 qd_2) - 2qd_2) \sinh(qd_1)^2\right]. \label{eq:Db}
\end{align}
For symmetric gates $d_1 = d_2 = d_g$, Eq.~\eqref{eq:Db} can be simplified as 
$D_b(q d_g) = [\tanh(q d_g) - q d_g \mathrm{sech}(q d_g)^2]$.
Introducing the effective impurity concentration $n_b = N_b / 2k_F$, the background impurity scattering rate [cf. Eqs.~\eqref{eq:f_no_gate} and~\eqref{eq:f0_no_gate}] $\tau_b^{-1} = \tau_0^{-1}(n_b) f_b$ can be rewritten as
\begin{align}
    &f_b(s,t,a,b_1,b_2) = \nonumber \\
    &\frac{2}{\pi} \int \limits_0^1 \frac{2 x (1-x^2)^{-1/2} D_b(x\sqrt{a} b_1,x\sqrt{a} b_2) dx}{\qty[\frac{x\sqrt{a}}{s} + \tilde{\Pi}(x\sqrt{a}, t) D(0,x\sqrt{a} b_1,x\sqrt{a} b_2)]^2},\label{eq:background_f}
\end{align}
and 
\begin{align}\label{eq:background_f0}
    f_{0b}(s,b_1,b_2) = \frac{2}{\pi} \int \limits_0^1 \frac{2 x D_b(x b_1,x b_2) dx}{\sqrt{1-x^2} \qty[\frac{x}{s} + D(0,x b_1,x b_2)]^2}.
\end{align}
In the limits of $b_1, b_2 \gg 1$ where the gates are far away, $D_b(x b_1,x b_2) \approx 1$ and $D(0,x b_1,x b_2) \approx 1$, such that Eqs.~\eqref{eq:background_f} and~\eqref{eq:background_f0} recover the results without gate screening Eqs.~\eqref{eq:f_b_no_gate} and~\eqref{eq:f_0b_no_gate}.
On the other hand, in the limit of low density $k_F d_1, k_F d_2 \ll 1$, we have 
\begin{align}\label{eq:D_b_small_b}
    D_b(x b_1,x b_2) \approx \frac{4 b_1^2 b_2^2 x^3}{3(b_1 + b_2)}, \, b_1, b_2 \ll 1.
\end{align}
Substitute Eq.~\eqref{eq:D_b_small_b} into \eqref{eq:background_f0} we obtain
\begin{align}
    f_{0b}(s,b_1,b_2) \approx \frac{4 b_1^2 b_2^2}{3(b_1 + b_2)\qty(s^{-1} + \frac{2b_1 b_2}{b_1 + b_2})^2}, \, b_1, b_2 \ll 1.
\end{align}
As a result, BIs behave as in-plane charged impurities at low densities $k_F d_1, k_F d_2 \ll 1$, where the BI scattering rate $\tau_{\mathrm{BI}}^{-1}$ is independent of $k_F$
\begin{align}\label{eq:tau_BI_small_kF}
    \frac{1}{\tau_{\mathrm{BI}}} = \qty(\frac{2}{g})^2 \frac{\pi^2\hbar}{m} \frac{4 N_b d_1^2 d_2^2}{3(d_1 + d_2)\qty(q_{TF}^{-1} + \frac{2d_1 d_2}{d_1 + d_2})^2}.
\end{align}
The zero-temperature mobility contributed by remote and background charged impurity scattering is given by $\mu_{\mathrm{RI}} = e \tau_0(n_r) / m f_0$ and $\mu_{\mathrm{BI}} = e \tau_0(n_b) / m f_{0b}$ respectively.
Using the Matthiessen addition rule $\mu^{-1} = \mu_{\mathrm{RI}}^{-1} + \mu_{\mathrm{BI}}^{-1} + \mu_{\mathrm{PD}}^{-1}$, we obtain the total mobility at zero temperature as shown by the solid black curves in Figs.~\ref{fig:muT_n} (c) and (f).
The corresponding results of the mobility and resistivity as a function of temperature are shown in Figs.~\ref{fig:muT_n} (b), (e) and Fig.~\ref{fig:rho_T} (b), respectively.
%, where curves of different colors correspond to different densities.
%The two scattering mechanisms, remote impurity scattering dominated at low densities and the scattering from short-range charge-neutral atomic point defects dominated at high densities, which explains the non-monotonic density dependence of the mobility observed in the experiments. 
\section{Remote and background charged impurity scattering at low densities}
\label{sec:RI_BI}
In this section, we discuss the low-density and low-temperature ($T=0$) mobility dominated by charged impurity scattering, and we ignore all other scatterers such as atomic point defects unless explicitly mentioned. 
Note that [see, e.g., Fig.~\ref{fig:mobility_nh}] the low (high) density mobility are determined by remote (background) impurity scattering, which is a generic result for 2D transport in the presence of both background and remote impurities.
This discussion is directly related to the Columbia sample~\cite{CDean_WSe2:2023}, where the low-density data are accessible because of more transparent contacts.
Remote impurities are always important for sufficiently low density where $k_F d\ll 1$ (i.e., $n \lesssim d^{-2}$).
This can be seen in the exponential suppression factor $e^{-qd}$ appearing in the RI Coulomb potential $U(q,d)$, which becomes ineffective at low densities $k_F d \ll 1$.
As a result, RIs behave similarly to in-plane charged impurities at low densities, where $\tau_{\mathrm{RI}}^{-1}$ saturates and becomes independent of $n$. 
If those RIs are close to the double gates, $\tau_{\mathrm{RI}}^{-1}$ is modified by a geometrical factor given by Eq.~\eqref{eq:rate_RI_near_gate}.
Similarly, BIs also behave as in-plane charged impurities at low densities, since BIs can be viewed as many layers of RIs separated from the 2D channel by difference distances. 
In the presence of double gates, $\tau_{\mathrm{BI}}^{-1}$ for $k_F d \ll 1$ is given by Eq.~\eqref{eq:tau_BI_small_kF}.
In our theory, at such low densities, depending on the details (for example, the impurity concentrations $n_r$ and $N_b$ and the hBN thickness $d_1$ and $d_2$ that determines the geometric factors), one may get RI dominating over BI scattering since, for $k_F d\ll 1$, RI behaves the same as BI.  
For example, if very large values of $n_r$ are used, obviously they will eventually dominate in the limit of very low density when $k_F d\ll 1$ even for the remote value of $d$.
On the other hand, to obtain concrete estimates of $n_r$ and $N_b$ and to determine which better fits the data, we need to use the difference between BI and RI scattering. 
At intermediate densities $k_F d\gg 1$, while the densities are still low enough such that point defect scattering is irrelevant, the RI and BI mobilities behave differently where $\mu_{\mathrm{RI}} \propto n^{1.5}$ and $\mu_{\mathrm{BI}} \propto n^{0.5}$~\cite{Hwang_density_scaling:2013}.
It is this distinction between $\mu_{\mathrm{RI}}$ and $\mu_{\mathrm{BI}}$ that determines which scattering mechanism is more suitable for fitting the low-density (more precisely, the intermediate-density) data.
The point to emphasize is that the distinction between ``remote'' and ``background'' is meaningful only when the two types of scatterer have different values of the dimensionless ``distance'' parameter $k_Fd$ with $k_Fd\gg1$ being remote and $k_Fd\ll1$ being background (or ``near'') scattering.  
Since $k_F \sim n^{1/2}$, for a sufficiently low carrier density $n$, the condition $k_F d\gg1$ cannot be satisfied for any scatterer and all scattering becomes scattering by near impurities.  
In such a situation, if there is a large amount of remote impurities, the remote scattering always dominates mobility, making the low-density mobility sensitive to the concentration of the remote scatterers.  
By contrast, the high-density mobility is dominated typically by background scattering since for large $n$, the remote scattering is strongly suppressed by virtue of the $2k_F d$ factor.  
The key point is that all scattering becomes important (i.e., both RI and BI mobility are independent of $n$) for sufficiently low carrier density.  
The easiest way to enhance the low-density mobility is therefore to set the separation of the remote scatterers large, which for the TMD layers under consideration can be done by increasing the hBN thickness.  
We predict that increasing the separation of the remote scatterers from the TMD layer will substantially enhance the low-density mobility without much affecting the high-density mobility.

For the Columbia sample~\cite{CDean_WSe2:2023}, to demonstrate that RI is the dominant scattering source at low densities, we show the comparison between BI and RI in Figs.~\ref{fig:mobility_nh} (a) and (b). 
Solid (dashed) curves represent the fits where the RI (BI) scattering dominates at low densities. 
As seen clearly in Fig.~\ref{fig:mobility_nh} (b) plotted in the log-log scale, the BI mobility follows a power law $\mu_{\mathrm{BI}} \propto n_h^{0.5}$, while the RI mobility follows $\mu_{\mathrm{RI}}\propto n_h^{1.5}$ for $k_F d \gg 1$~\cite{Hwang_density_scaling:2013}. 
Since the low-density mobility data are proportional to $n_h^{\alpha}$ with $\alpha\approx 1.5$, this indicates that RI (with large value of $n_r$ and small value of $N_b$) fits the data better.
In addition, the RI dominant fit starts to deviate from the data at a density $n_h\lesssim 5\times 10^{11}$ cm$^{-2}$, while the BI dominant fit starts to deviate from the data at a higher density $n_h\lesssim 1\times 10^{12}$ cm$^{-2}$. This also makes the RI dominant fit better compared to BI.
We see that at sufficiently low densities $n_h\lesssim 1\times 10^{11}$ cm$^{-2}$, the theoretical RI and BI mobility curves acquire smaller slopes and eventually saturate, as anticipated (i.e., RI and BI behave the same, and the distinction between them has disappeared as $k_Fd\ll 1$ for all scatterers).
We should emphasize that, since the Boltzmann transport theory becomes increasingly unreliable as one approaches the low-density metal-insulator transition, it is reasonable that both the RI and BI results deviate from the experimental data at sufficiently low carrier density.

Next, we comment on the relatively large number of $n_r = 3\times 10^{12}$ cm$^{-2}$ obtained from the RI dominant best-fit result in Fig.~\ref{fig:muT_n} (c).
This number is obtained specifically for the case where the RIs are separated from the gate by a distance $d_0 = 1$ nm, and it changes sensitively when $d_0$ changes. 
For example, we show the mobility results for the same impurity concentrations, but with $d_0 = 2$ nm [twice larger than the value $d_0=1$ nm used in Fig.~\ref{fig:muT_n} (c)], plotted as dashed lines in Figs.~\ref{fig:mobility_nh} (c) and (d). 
The RI mobility drops by a factor of $\sim 5$, because the gate screening effect is weaker for larger $d_0$. 
This can be seen from the fact that the dipole length made by the RI and its image charge is proportional to $d_0$, and the scattering rate is proportional to $d_0^2$ [cf. Eq.~\eqref{eq:rate_RI_near_gate}]. 
This implies that the RI concentration $n_r$ should be reduced by a factor $\sim 5$ to fit the data. 
As a result, $n_r$ strongly depends on the choice of $d_0$, but $d_0$ is an unknown parameter.
In this sense, the number $n_r = 3\times 10^{12}$ cm$^{-2}$ is only our best estimate based on the empirical guess that if there are RIs then they should be located near the interface. 
In experimental samples, the only remote interface is the hBN/graphite-gate interface, and $d_0 \sim 1$ nm is not an unreasonable number.

We suggest a way to improve the peak mobility in experiments if the low-density mobility is indeed limited by RI scattering near the hBN/graphite-gate interface.
Given that RI scattering is sensitive to $d$, as seen by $\mu_{\mathrm{RI}} \propto (k_F d)^3$ [cf. Eq.~\eqref{eq:rate_RI_near_gate}], the low-density mobility can be enhanced by increasing $d$.
For example, we theoretically predict the mobility by increasing the top hBN thickness to $d_1 = 25$ nm (cf. $d_1 = 11$ nm in the original experiment~\cite{CDean_WSe2:2023}), plotted as dashed curves in Figs.~\ref{fig:mobility_nh} (e) and (f), while keeping all other parameters the same in calculations.
We find that the peak mobility reaches $10^{5}$ cm$^2$/Vs, because the low-density RI mobility increases by more than an order of magnitude and becomes comparable to the BI mobility. 
Note that this also obviously shifts the occurrence of the peak mobility to a lower carrier density $n_h\sim 3\times 10^{11}$ cm$^{-2}$ since remote scattering is being suppressed by increasing the hBN thickness.
(On the other hand, we note that the contact resistance increases strongly around $\lesssim1.7\times 10^{11}$ cm$^{-2}$ in experiments, which could somehow affect the observation of the peak mobility around the same density.)
In conclusion, by increasing the hBN thickness $d$, RI scattering is suppressed and the low-density mobility increases.  
At sufficiently large $d$, BI scattering becomes the dominant scattering source at low carrier densities.

\section{phonon scattering}
\label{sec:phonon}
In this section, we briefly discuss and compute the phonon scattering rate within the deformation potential approximation~\cite{Kaasbjerg:2012,Jin_Zhenghe:2014,Zhang_Wanli:2016,Hwang:2019}.
The acoustic phonon scattering rate reads~\cite{Fivaz:1967,Kawamura:1992}
\begin{align}
    \frac{1}{\tau_{\mathrm{A}}} = \frac{m D^2}{\hbar^3 \rho_m v_s^2} T.
\end{align}
The linear-in-$T$ resistivity regime crosses over to the Bloch-Gr\"{u}neisen (BG) regime $\propto (T/T_{BG})^4$ at low temperatures $T\lesssim T_{\mathrm{BG}}/3$~\cite{Lavasani:2019,Hwang:2019,Poduval_Kondo:2022}, where $T_{\mathrm{BG}} = 2 \hbar v_s k_F$ and $v_s$ is the sound velocity.
As a crude estimate, the mobility of WSe$_2$ monolayer at room temperature $T=$ 300 K assuming only longitudinal acoustic phonon scattering is $\mu_{\mathrm{LA}} = 1.3 \times 10^3$ cm$^{2}$/Vs, where we use the 2D mass density $\rho_m = 6.0 \times 10^{-7}$ g/cm$^{2}$, the electron-phonon coupling $D = 3.78$ eV~\cite{Zhang_Wanli:2016}, and the sound velocity $v_s = 3.3 \times 10^5$ cm/s~\cite{Jin_Zhenghe:2014}.
This number of $\mu_{\mathrm{LA}}$ agrees reasonably well with the experimental data as shown in Fig.~\ref{fig:mu_T}.
The optical phonon scattering rate for the zero-order deformation potential scattering reads~\cite{Kaasbjerg:2012}
\begin{align}
    \frac{1}{\tau_{\mathrm{O}}(\epsilon)} = \frac{m D_0^2}{2 \hbar^2 \rho_m \omega_{\lambda}} [N_{\lambda} + (N_{\lambda}+1) \Theta(\epsilon - \hbar \omega_{\lambda})],
\end{align}
where $D_0$ is the zero order deformation potential and $\omega_{\lambda}$ is the energy of the $\lambda$-th optical phonon mode with an average occupation number $N_{\lambda} = (e^{\hbar\omega_{\lambda}/T} - 1)^{-1}$. 
The two terms inside the bracket represent the absorption $\propto N_{\lambda}$ and emission $\propto (N_{\lambda} + 1)$ of phonons, respectively. 
The Heaviside theta function ensures that only electrons with sufficiently large energy can emit a phonon.
The dominant temperature dependence of the optical phonon scattering comes from the phonon occupation number even after doing the energy average, such that $\tau_{\mathrm{O}} \propto N_{\lambda}^{-1}$.
The effective power-law temperature dependence of the mobility $\mu \propto \tau_{\mathrm{O}} \propto T^{-\gamma}$ can be obtained by $\gamma = d \ln N_{\lambda}/d \ln T$, and we get $\gamma = 1$ at $T \gg \hbar\omega_{\lambda}$ while $\gamma = 1.58$ at $T \ll \hbar\omega_{\lambda}$.
Since the energy of typical optical phonon modes in WSe$_2$ is $\sim 300$ K~\cite{Jin_Zhenghe:2014}, we get $\gamma \approx 1.2$ at room temperature.
This analysis is in reasonable agreement with the fitting curves to experimental data with a slope ranged from $\gamma = 1$ to $\gamma = 1.5$ as shown in Fig.~\ref{fig:mu_T}.

\section{Conclusion}
\label{sec:conclusion}
In summary, our research focuses on examining electronic transport at low temperatures in monolayer TMDs under realistic disorder conditions inspired by recent experiments~\cite{CDean_WSe2:2023,PKim:2023,sung2023observation}. 
We analyze various aspects of the experimental resistivity data, such as the 2D MIT behavior, the Wigner crystallization, and the temperature and density dependence of mobility and resistivity in the metallic phase. 
The linear-in-$T$ metallic resistivity experimentally observed at low temperatures is explained by the temperature-dependent Friedel oscillations associated with screened charged impurities. 
We explore the possibility that this Coulomb disorder may also be responsible for triggering the MIT as either a disorder-driven quantum Anderson localization transition or a classical percolation transition through the long-range disorder potential landscape. 
The theoretically predicted critical density for disorder-induced MIT is lower than the experimental critical density, but only by a factor of $\sim$2 (which is not unreasonable given the approximations involved in the theory and the many unknown experimental details).
This suggests that the observed 2D MIT behavior in the high $r_s \sim 30$ regime likely results from the interplay between disorder and interaction-driven Wigner crystal physics.
Nevertheless, because of their large effective mass and low dielectric constant, TMDs have a substantial melting temperature and high critical density for the Wigner crystal as predicted by the mean-field theory, making them one of the most suitable materials for investigating Wigner crystal physics.
We briefly discuss the weak localization effect, although it is not observed in the experimental data.
In the diffusive limit $T\tau/\hbar \ll 1$, the conductivity acquires a logarithmic correction whose numerical prefactor depends on the details of the weak localization effect and the electron-electron interaction~\cite{Altshuler_Aronov:1985,Finkelstein:1983,Finkelstein:1984,Castellani:1984,Castellani:1984b,Castellani:1998,Zala_Aleiner:2001}.
If the phase coherent time $\tau_\varphi$ is determined by inelastic electron collisions, then both the weak localization corrections to the Boltzmann conductivity and the corrections originating from interaction are proportional to $\ln T$ and will differ only in numerical coefficients~\cite{Altshuler_Aronov:1985,Zala_Aleiner:2001}.
However, this $\ln T$ insulating behavior is difficult to observe due to the competing linear-in-$T$ metallic Boltzmann resistivity, which may only become apparent at extremely low temperatures outside the experimental temperature range~\cite{Tracy:2014}.
There are many approximations made in our theory in order to calculate actual numerical results for a comparison with experiments.  
We use the Boltzmann transport theory within the leading-order relaxation time approximation, which becomes increasingly inaccurate as the carrier density is lowered toward the MIT, but the theory never becomes invalid until there is an actual phase transition (e.g. at the MIT).  
However, the theory should be quantitatively accurate deep into the metallic phase, and indeed we get quantitative agreement with the experimental resistivity at high carrier density, but our estimate of the critical density for the MIT is off by a factor of 2.  
Our screening of the Coulomb disorder uses the random phase approximation, which is also exact only at high densities, becoming inaccurate at lower densities.  
Going beyond these approximations is difficult, if not impossible, and therefore, it is satisfying that we get a reasonable agreement with the experimental transport data over a broad density range with our theory becoming quantitatively inaccurate at low densities as expected.  
We also find that associating the very low density (i.e. large $r_s$) observation of the MIT does not necessarily imply the observation of WC since disorder-induced localization effects are also strongly enhanced at lower densities-- this point has already been emphasized earlier in the literature focusing on 2D semiconductor systems~\cite{Ahn_Anderson_Wigner:2023}. 
In fact, the reported observation of the lowest critical density (highest critical $r_s$) MIT was interpreted as a percolation transition rather than a WC transition, although the critical $r_s$ for the MIT in that 2D hole p-GaAs system was $\gtrsim 50$, much higher than the putative theoretically accepted value ($\sim30$) for the 2D WC transition~\cite{Manfra:2007}. 
We believe that low-density 2D systems manifest a complex interplay of disorder and correlation effects, which cannot be easily described as a simple WC transition, even when the critical $r_s$ is as large as 30 (or higher), provided that the AIR condition predicts a localization MIT also near this high critical $r_s$. 
This becomes obvious once we take into account the fact that the mean free path for the localization MIT according to the AIR criterion is $l\sim 1/k_F$, which, by definition, is comparable to the WC lattice constant, implying that any such WC necessarily manifests periodic coherence only over one (or at most, a few) lattice constants. 
Such a strongly disordered WC could also be construed equivalently as an Anderson insulator with short-range correlations induced by interactions.

\begin{acknowledgments}
The authors thank C.R. Dean, P. Kim, H. Park, A.Y. Joe, E. Demler, I. Esterlis, J.D. Sau, Q. Shi, J. Pack, and J. Sung for useful discussions.
This work is supported by the Laboratory for Physical Sciences.
\end{acknowledgments}

\medskip
%\bibliographystyle{apsrev4-1}
%\bibliography{reference.bib}

\begin{thebibliography}{130}
\bibitem [{bbg()}]{bbg_dielectric}%
  \BibitemOpen
  \href@noop {} {}\bibinfo {note} {We use the average dielectric constant between hBN and vacuum $\kappa \sim 3$ for the BLG/hBN device.}\BibitemShut {Stop}%
\end{thebibliography}
%\medskip

\end{document}